# Effects of macroH2A and H2A.Z on nucleosome structure and dynamics as elucidated by molecular dynamics simulations

Samuel Bowerman and Jeff Wereszczynski[*]

*Department of Physics and Center for Molecular Study of Condensed Soft Matter, Illinois Institute of Technology, 3440 S Dearborn St., Chicago, IL 60616, USA*

E-mail: jwereszc@iit.edu

## Abstract

Eukaryotes tune the transcriptional activity of their genome by altering the nucleosome core particle through multiple chemical processes. In particular, replacement of the canonical H2A histone with the variants macroH2A and H2A.Z has been shown to affect DNA accessibility and nucleosome stability; however, the processes by which this occurs remain poorly understood. Here, we elucidate the molecular mechanisms of these variants with an extensive molecular dynamics study of the canonical nucleosome along with three variant-containing structures: H2A.Z, macroH2A, and an H2A mutant with macroH2A-like L1 loops. Simulation results show that variant L1 loops play a pivotal role in stabilizing DNA binding to the octamer through direct interactions, core structural rearrangements, and altered allosteric networks in the nucleosome. All variants influence dynamics; however, macroH2A-like systems have the largest effect on energetics. In addition, we provide a comprehensive analysis of allosteric networks in the nucleosome and demonstrate that variants take advantage of stronger interactions between L1 loops to propagate dynamics throughout the complex. Furthermore, we show that post-translational modifications are enriched at key locations in these networks. Taken together, these results provide new insights into the relationship between the structure, dynamics, and function of the nucleosome core particle and chromatin fibers, and how they are influenced by chromatin remodelling factors.

---

[*]To whom correspondence should be addressed

# Introduction

Eukaryotes package their genetic code in highly ordered chromatin fibers. The fundamental unit of these structures is the nucleosome core particle (NCP), a complex of ∼147 basepairs of DNA that are wrapped around eight histone proteins (Figure 1).[1] Although they have minimal sequence homology, each core histone has a structural motif of an N-terminal tail, three α-helices connected by two loops (α1-L1-α2-L2-α3), and a C-terminal tail.[1,2] In the assembled NCP, histones are structurally divided into a (H3-H4)$_2$ tetramer that is positioned between two H2A-H2B dimers. The only location of inter-dimer interactions is at the base of the NCP which is formed by the H2A L1 loops, whereas each dimer has two interfaces with the tetramer: the H2A "docking domain" (DD) and the H3-H4 "four helix bundle."[1–4]

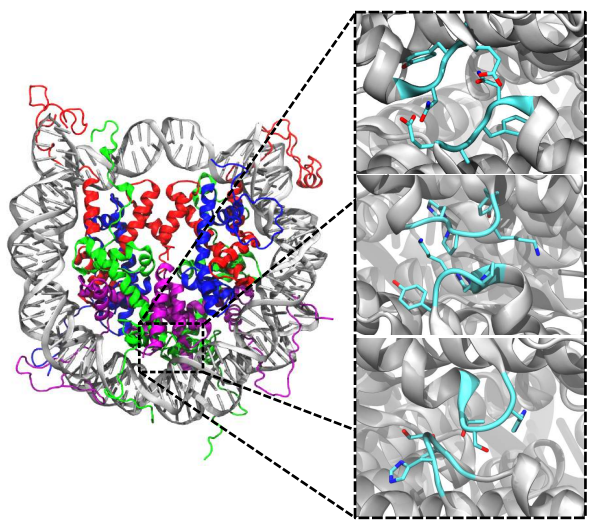

**Figure 1:** (left) The nucleosome core particle viewed down the DNA superhelical axis. Coordinates are taken from the final snapshot of the first canonical nucleosome simulation. Color code: H3 (red), H4 (blue), H2A (green), H2B (purple), and DNA (grey). (right) The structures of the three L1 loop sequences considered in this study: canonical (top), macroH2A (middle), and H2A.Z (bottom). The canonical loops possess a net negative charge resulting from Glu41, while the macroH2A loops possess a net positive charge from Lys40. The L1 loops of H2A.Z are uncharged, but both macroH2A and H2A.Z loops introduce a larger hydrophobic volume than the canonical.

Cells regulate chromatin stability and DNA accessibility by changing the biochemical properties of the NCP.[5–8] One of the primary chromatin remodeling mechanisms is the replacement of H2A or H3 histones with "histone variants."[9–15] These variants have a similar structure and sequence to the canonical histones, however they diverge at key locations that affect inter-histone and DNA-histone contacts. These differences alter the structure and stability of the NCP and are therefore implicated in modulating transcriptional activity. For example, the H2A variant macroH2A exists in large populations in the inactive X chromosome of females but is sparse in active genes.[11,16,17] In contrast, the H2A.Z variant has been linked to both transcriptional activation and repression and is enhanced in regulatory regions of the genome such as promoters and enhancers.[18,19]

Histone variants influence chromatin through diverse mechanisms and structure/function relationships. macroH2A is unique among variants in that it possesses multiple domains, including the histone domain, a 38 residue linker sequence, and a large "macro-domain."[3,20] On its own, the histone domain is sufficient for reducing transcriptional activity *in vivo* and increasing the stability of the nucleosome complex, even though the crystal structure of an NCP containing this domain shows that variant incorporation causes only minor NCP rearrangements.[11,21] The primary sequence is ∼65% identical to canonical H2A and differs largely from H2A in two important regions: the L1 loops and the docking domain. The canonical $^{38}$NYAE$^{41}$ L1 loop possesses a net negative charge, while in contrast the macroH2A L1 $^{38}$HPKY$^{41}$ sequence has a net positive charge and an increased hydrophobicity. Substitutions of the L1 loops in canonical H2A with a macroH2A sequence (the "L1-Mutant") creates nucleosomes with *in vitro* stabilities and *in vivo* enrichments that are nearly identical to NCPs containing the complete macroH2A histone domain.[11,21] Therefore, the L1 loops appear to be pivotal in dictating macroH2A's abilities to affect intra-nucleosomal functions. Meanwhile, changes to the docking domains show little effect on *in vitro* stability, but increase *in vivo* enrichment.[11,21]

The role and mechanisms of the H2A.Z variant remains less well defined, with some experiments showing that H2A.Z increases NCP stability while others have found that it destabilizes the system. Similar to macroH2A, a comparison of the H2A.Z and canonical containing NCP crystal structures show nearly identical overall conformations with the exception of two features.[4] First, the structure of the L1 loops is altered, resulting in increased contacts between the two H2A/H2B dimers which likely helps to stabilize the histone octamer. Second, H2A.Z has fewer hydrogen bonds between the docking domain and H3, which could destabilize the dimer/tetramer interface. This combination of stabilizing one area of the NCP while destabilizing another may account for the disparate experimental results and the multiple functions H2A.Z appears to have.[22]

Experiments have revealed a wealth of information about how histone variants affect NCP and chromatin function, yet several questions still remain. For example: how do seemingly minor structural rearrangments affect the stability of the nucleosome? To what extent do changes in the L1 loops propagate through the complex? Do variants influence NCP function through only structural means, or do they take advantage of altered dynamics as well? To address these problems, we have performed extensive molecular dynamics (MD) simulations of four complete NCP systems that include: 1) canonical H2A histones, 2) the macroH2A histone fold domains, 3) "L1-Mutant" H2A histones, and 4) H2A.Z histones. Our results indicate that different sequences in the L1 loops perturb the dynamic and energetic properties in this region of variant containing NCPs. These effects propagate throughout the complex and create subtle, yet important, rearrangements that alter the NCP structure and dynamics through both direct effects and modified allosteric networks. This allows histone variants to influence both the global dynamics and energetics of the NCP, and likely contributes to large-scale structural changes such as DNA breathing and nucleosome opening, as well as inter-NCP interactions in chromatin fibers.[23,24]

# Materials and Methods

## System and Simulation Details

Simulations of the canonical, macroH2A, and H2A.Z containing nucleosomes were initialized from their crystal structures (PDB: 1KX5, 1U35, and 1F66, respectively).[2–4] The L1-Mutant structure was formed using the crystal structure of 1KX5, with the H2A L1 loops mutated from the canonical $^{38}$NYAE$^{41}$ to the macroH2A $^{38}$HPKY$^{41}$ sequence. The systems were neutralized and solvated in a 10 Å TIP3P box of 150mM NaCl, creating systems of approximately 250,000 atoms. Each system was simulated three times (see supplemental material for more details). The simulations were done in the NAMD engine (v2.9) using the AMBER12SB fixed point-charge forcefield.[25,26] Monovalent ions were modeled according to Joung and Cheatham.[27] Production simulations were done in the NPT ensemble using standard techniques.[28–31] Coordinates were stored every 2 ps. Visualizations were made using VMD and PyMOL.[32–34]

### Allosteric Pathways Calculations

Allosteric effects were computed with multiple techniques (for specific details, see supplemental materials). Per-residue differences in dynamics were determined by calculating the Kullback-Leibler divergence of dihedral angle populations.[35] For these calculations, the canonical populations were used as a reference set. Spurious results were filtered using bootstrapping techniques. Residue-residue correlations were calculated by utilizing the "largest mutual information" method.[36,37] Residue contacts were determined to be when protein $C\alpha$ or nucleic C1' atoms were within 10 Å in at least 70% of the configurations.[38] The mapping of allosteric networks was conducted using the Weighted Implementation of Sub-optimal Pathways approach.[39] The edge-betweenness centrality of residues in the optimal networks were calculated with the NetworkX Python package, with the significance determined by a hypergeometric distribution (see supplemental materials).[40–42]

### Interaction Energies

Interaction energies between the L1 loops were calculated using cpptraj.[43] A cutoff distance of 15 Å was used for both van der Waals and electrostatic interactions. Because the L1 loops interact with both protein and DNA, an intermediate dielectric value of 5 was considered ($\varepsilon_{DNA} = 8$, $\varepsilon_{protein} = 4$).[44,45] Hydrogen bonds were defined by a separation distance of 3.5 Å and an angle of 30°.

DNA binding and complex assembly energies were calculated using the MMPBSA.py function of AMBERTOOLS (v.14).[46] The level of theory was restricted to the Generalized Born Implicit Solvent (igb=5, radii=mbondi2).[47] Coordinates from every 100ps of production simulation were used. The coordinates for the protein constituents were extracted from the nucleosome simulations, but the unbound DNA coordinates were taken from a separate simulation of 147bp of linear B-form DNA in a 150 mM NaCl environment. Error bars in the energies and all other measures are defined by the standard error of the mean, where the number of independent points was determined by the statistical inefficiency of the data set, as computed with the PyMBAR package.[48]

## Results

We performed three independent 250 ns MD simulations for four complete NCP systems: 1) canonical, 2) macroH2A histone fold, 3) "L1-Mutant," and 4) H2A.Z containing NCPs. In each set of simulations, approximately 50 ns was required for the root-mean square deviations (RMSDs) of the complexes to stabilize (Figures S2-S5) and for the tails to collapse from their initial elongated states. These results are consistent with previous MD simulations of the canonical NCP which demonstrated that the overall complex is stable on the hundreds of ns timescale and that the histone tails form strong interactions with the nucleosomal DNA.[23,49,50] Comparisons between the canonical and variant systems demonstrate that variants have both subtle and large-scale effects on the structure and dynamics of the L1 loops, DNA-histone interactions, and allosteric networks throughout the NCP.

## Altered Dynamics of L1 Loops

Modifications of the L1 loop sequences in histone variants alter their dynamics and energetics. In the canonical NCP simulations, an average of 0.5 hydrogen bonds were formed between the L1 loops, primarily between the carboxamide nitrogen of asparagine and the carboxylate group of the symmetric glutamate. This is consistent in the L1-mutant (∼0.4) but is reduced to ∼0.2 in macroH2A. In these two systems, the most prevalent hydrogen bonds were formed between the phenol oxygen of tyrosine and the lone pair of the symmetric histidine. In the H2A.Z simulations, L1-L1 hydrogen bonds were almost nonexistent (Table S1).

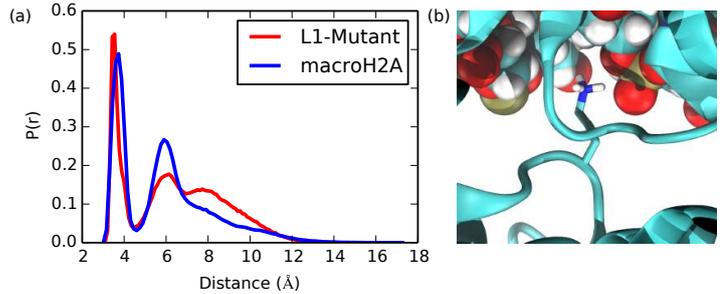

**Figure 2:** (a) Distance populations for Lys$^{40}$ to DNA phosphate show an interaction that is unique to macroH2A-like L1 loops. (b) A representative configuration of the Lys sidechain stretching across the molecule to interact with the dimer's non-associated DNA. This orientation sterically hinders the symmetric loop from forming a similar interaction. This interaction contributes significantly to stabilizing DNA-octamer binding in the macroH2A and L1-Mutant systems.

Although hydrogen bonds form most frequently in the canonical loops, the net L1 loop interaction energies are more favorable in the variants (Table 1). The close proximity of the negatively charged glutamates in the canonical NCP creates a disfavored electrostatic interaction. However, the L1-Mutant and macroH2A systems avoid an analogous situation through Lys-DNA interactions, which separates the like-charges and creates a more favorable electrostatic configuration. Meanwhile, the lack of charge in the H2A.Z L1 Loops also creates a more favorable electrostatic interaction than the canonical NCP. In addition, the L1 loop rearrangement in the macroH2A and L1-Mutant systems is further stabilized by van der

**Table 1:** Net interaction energies between H2A L1 loops show that macroH2A and H2A.Z loops have more favorable interactions than canonical L1 loops. ΔU is defined as the difference in energies between each variant and the canonical system. Negative values show favorability in the variants. All values are reported in kcal/mol.

| System | $U_{elect}$ | $U_{vdW}$ | $U_{tot}$ | $\Delta U_{elect}$ | $\Delta U_{vdW}$ | $\Delta U_{tot}$ |
|---|---|---|---|---|---|---|
| Canonical NCP | 6.3 ± 2.0 | -9.4 ± 0.8 | -3.1 ± 2.2 | —— | —— | —— |
| L1 Mutant | -0.9 ± 0.4 | -13.4 ± 0.8 | -14.3 ± 0.9 | -7.2 ± 2.0 | -4.0 ± 0.8 | -11.2 ± 2.4 |
| macroH2A NCP | 0.4 ± 1.1 | -12.1 ± 0.9 | -11.7 ± 1.0 | -5.9 ± 2.3 | -2.7 ± 0.9 | -8.6 ± 2.3 |
| H2A.Z NCP | -0.6 ± 0.1 | -6.9 ± 0.6 | -7.5 ± 0.6 | -6.9 ± 2.0 | 2.4 ± 1.0 | -4.4 ± 2.3 |

Waals interactions. In total, the interaction energies of the L1 loops in the L1-Mutant and macroH2A structures are substantially favored over those of the canonical nucleosome, with respective $\Delta U_{total}$ values of -11.2 ± 2.4 kcal/mol and -8.6 ± 2.3 kcal/mol. The H2A.Z L1 loop conformations are also more favorable than in the canonical system ($\Delta U_{total}$ = -4.4 ± 2.3 kcal/mol).

The different net charges of the L1 loops influence their interactions with the nucleosomal DNA. In the canonical L1 loops, the negative charge located on Glu41 causes a repulsive force to the negatively charged DNA. However, in the macroH2A and L1-Mutant systems, Lys40 introduces a positive charge into the loop which forms a salt bridge with the DNA basepair across the axis of symmetry (Figure 2). The lysine forming this salt bridge sterically hinders the symmetric lysine residue from doing the same, so the interaction exists in only one dimer. The non-interacting lysine is primarily exposed to solvent while intermittently forming a hydrogen bond with a neighboring histidine. Since the L1 loops of H2A.Z are uncharged, they are not capable of forming similar interactions, and therefore did not make any direct contacts with the DNA. Taken with the results of the L1-L1 loop dynamics, we observe that the macroH2A-like loop sequences stabilize both protein-protein and protein-DNA interactions when compared to both the canonical and H2A.Z histones.

## Variant Presence Alters Dimer Orientations

Reorganization of the L1-loops creates perturbations that affect the dimer orientations in the NCP. For example, the H2A α2 helix extends across the dimer, with its N-terminal (the "base") at the L1-interface and its C-terminal (the "top") solvent-exposed on the far side of the molecule (Figure 3). Simulation analysis showed that the canonical system exhibited a separation of 67.1 ± 0.2 Å between the tops, and 34.8 ± 0.1 Å between the bottoms of the H2A α2 helices. In contrast, in each of the variant NCPs there is a "bulging" motion in which the base separation is increased to ~36.5 Å while the top separation is decreased to ~66.1 Å. Although these changes in orientation are only on the order of an Å, a t-test indicated that they are all extremely statistically significant (Figure 3).

This subtle re-orientation of the dimers alters histone-DNA hydrogen bonding. For example, the guanidine group of H2A Arg29 forms a hydrogen bond with the phosphate group of the 23rd basepair of DNA in all systems. In the canonical NCP, this bond forms in 63% of the configurations, whereas in the L1-mutant it is formed 70% of the time. In addition, the frequency of hydrogen bonding between the 22nd basepair phosphate and the backbone amide of H2B Ser33 increases from 50% in the canonical system to 60% in the L1-Mutant structure. In H2A.Z, the hydrogen bonding at these locations increases drastically to 73% and 85%, respectively. Interestingly, the macroH2A nucleosome shows a decreased frequency of both of these interactions (52% and 30%, respectively). The reduced hydrogen bonding in the macroH2A nucleosome is likely a result of sequence deviations in the nearby H2A α1 helix (canonical: $^{30}$VH$^{31}$, H2A.Z: $^{30}$IH$^{31}$, macro: $^{30}$ML$^{31}$).

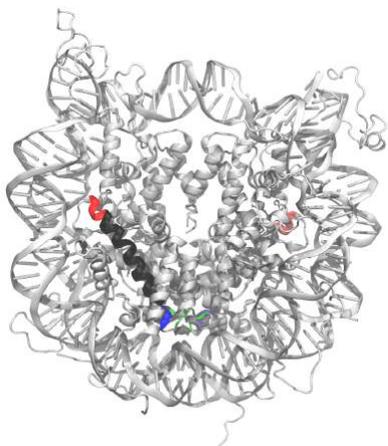

| System | top | bottom |
|---|---|---|
| Canonical NCP | 67.1 ± .2 | 34.8 ± .1 |
| L1-Mutant | 66.0 ± .2 | 36.8 ± .4 |
| p-value | .0003 | .0001 |
| macroH2A NCP | 66.2 ± .3 | 36.2 ± .1 |
| p-value | .0001 | .0001 |
| H2A.Z NCP | 66.0 ± .1 | 36.1 ± .1 |
| p-value | .0001 | .0001 |

**Figure 3:** Separation distances for H2A α2 helix locations show a "bulging" effect in histone variants. The helix is displayed in black while the helix "top" is highlighted in red and the "bottom" in blue. The L1-loops are shown in green for clarity. Shifts in mean separation are on the order of an Å, but the changes in populations are all incredibly significant.

The dimer realignment also affects the hydrogen bonding between protein constituents in the histone core. In the canonical NCP, an average of 14.8 hydrogen bonds are formed between a single dimer and the tetramer, which is in agreement with the ∼15 observed in the crystal structure. This increases to an average of 16.5 in the L1-Mutant. The macroH2A and H2A.Z nucleosomes display an average of 14.7 and 14.4 hydrogen bonds, which are substantially more frequent than the ∼8 observed in the initial configurations. Therefore, the variant dimer reorientation encourages the histones to form hydrogen bonds more frequently than in the crystal structures.

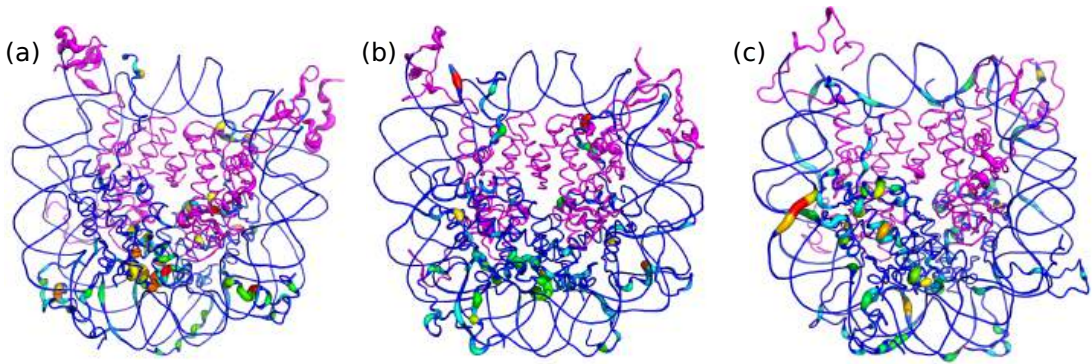

**Figure 4:** Kullback-Leibler Divergence of dihedral angles for the (a) L1-Mutant, (b) macroH2A, and (c) H2A.Z nucleosomes, using the canonical populations as a reference set. The dimers and DNA residues are represented in a rainbow spectrum, where divergence values increase from blue to red. The tetramer is shown in magenta, where the tube radius is wider for larger values. Significant divergences are observed both in the vicinity of and far from the L1-L1 interface.

## Histone H2A L1 Sequence Influences Dynamics Throughout the Nucleosome

The L1 loops not only influence dimer reorientation, but they also perturb the local dynamics of residues throughout the nucleosome. Calculations of the Kullback-Leibler (KL) divergence between the canonical and variant systems showed the expected disparity in dihedral sampling of residues within the L1 loop region (Figure 4). However, they also highlighted significant changes in the dynamics of residues that are distant from these loops. In both the L1-mutant and macroH2A systems, the dimer and tetramer constituents of the docking domains have statistically significant KL divergence values, indicating that their local dynamics are different in these systems relative to the canonical NCP. Although this was expected in the macroH2A and H2A.Z systems due to their sequence deviations, the L1-mutant is sequentially identical to the canonical system in these areas. Therefore, the observed difference in dynamics must be due to allosteric networks that are shifted by the L1 mutation. On the other hand, the dynamics in the histone core of H2A.Z only show small differences from the canonical system, most notably along the H2B $\alpha 2$ helix, while the largest divergences are in the DNA.

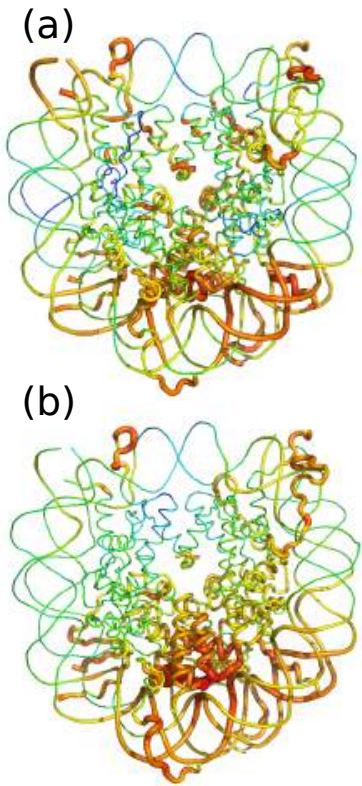

**Figure 5:** Average individual residue correlation with L1 loop residues for (a) canonical and (b) L1-Mutant nucleosomes. Thicker, redder residues are those with stronger average correlations with the L1 loop sequence. The L1-Mutant nucleosome shows increased correlations near the L1-L1 interface, as well as among H2B-H4 four helix bundle residues. Both systems display appreciable correlations between the L1 loops and docking domain residues in the dimer and tetramer. Histone tails were truncated to improve clarity.

The L1-mutant and macroH2A variants also have increased dynamical correlations between the H2A L1 loops and key portions of the NCP (Table 2 and Figure 5). The strengthened interactions in the L1 loops increase the L1-L1' correlations from 0.42 in the canonical system to >0.67 in each of the variants. In both the canonical and H2A.Z systems, the average correlation between the L1 loops and either docking domain (symmetric - DD, opposing dimer - DD') was 0.36-0.38. However, in the L1-Mutant the average L1-DD and L1-DD' correlations increased to 0.51 for both measurements, which were further increased to 0.56 and 0.60 in the macroH2A structure. Although the variants had an increased correlation between the L1 loops and DNA near the base of the nucleosome, the correlations between the L1 loops and DNA extremities were similar in all four systems.

## Variant Presence Alters Allosteric Pathways

The origins of the altered dynamics and correlations in NCPs with variants were probed by computing the optimal and suboptimal correlation pathways using with the Weighted Implementation of Suboptimal Paths algorithm[39]. Allosteric networks were calculated between the L1 loops and the DNA entry and exit sites, and the tetramer components of the docking domains for each system. The results revealed that not only are there several networks of dynamically coupled residues in the canonical NCP, but that these networks are both modified and strengthened by macroH2A, H2A.Z, and the L1-mutant. The shifts are due to both changes in the NCP hydrogen bonding networks from subtle repositioning of the H2A histones, as well as increased interactions of the L1 loops with one another and with the nucleosomal DNA.

In the L1-to-symmetric DNA end pathways, the canonical system utilizes three main routes for information transfer (Figure 6). In the first, networks primarily pass through neighboring H2B Ser$^{33}$-DNA and H2A $\alpha 1$ helix Arg$^{29}$-DNA hydrogen bond interactions and into the DNA, whereas in the second the networks enter the DNA through the H2A Arg$^{42}$-

**Table 2:** Average correlations between L1 loops and relevant regions of the nucleosome core particle for each system. The L1-L1' and L1-DD(') correlations are significantly stronger in the systems possessing the macroH2A L1 loops, while L1 correlations to the DNA extremities are unchanged. The associated docking domain is abbreviated as DD, and the docking domain of the opposing dimer is abbreviated as DD'.

| System | L1-L1' | L1-DD | L1-DD' | L1-DNA |
|---|---|---|---|---|
| Canonical NCP | 0.42 | 0.38 | 0.36 | 0.49 |
| L1 Mutant | 0.70 | 0.51 | 0.51 | 0.48 |
| macroH2A NCP | 0.73 | 0.56 | 0.60 | 0.48 |
| H2A.Z NCP | 0.67 | 0.38 | 0.38 | 0.44 |

DNA hydrogen bond near the intradimer interaction site. The third route for propagation extends along the H2A α2 helix, which passes dynamic information into the DNA basepairs via a Thr$^{76}$-DNA interaction. The pathways of H2A.Z are similar to the canonical but with more pathways accessing the H2A α1 Arg$^{29}$-DNA interaction than that of H2B Ser$^{33}$-DNA. In the L1-Mutant, the increased prevalence of the Arg$^{29}$-DNA hydrogen bond heavily biases information transfer through this network and increases the strength of this pathway. The decreased Arg$^{29}$-DNA interaction in the macroH2A nucleosome causes information to be transferred primarily via the H2A Arg$^{42}$-DNA hydrogen bond, with a significant number of pathways also traversing the H2A α2 helix.

The effects of L1-L1' communication transfer are most apparent in the networks between an L1 loop and the DNA end of the opposite symmetry. In the canonical nucleosome, there exist no pathways between L1 loops, therefore networks must pass through indirect routes that include the DNA and histone tails. However, in all of the variant structures information is readily exchanged between the L1 loops, allowing the pathways to immediately cross into the opposite symmetry dimer (Supplemental Figures S12-S15). Once information is passed into this dimer, it follows the typical pathways for L1-to-symmetric DNA end propagation. This results in allosteric networks that are not only stronger, but more direct in the histone variants.

Pathways between the L1 loop and docking domains in the same dimer are similar in all systems, but there is a large disparity in the pathways between L1 loops and the docking domain of the other dimer constituent. In the canonical NCP, the majority of paths pass from the L1 loops through the H2B α2 helix into the tetramer portion of the docking domain via the four helix bundle of H2B-H4. The L1-Mutant structure shows an increased number of contacts in this region, creating a more diverse set of pathways between bundle helices. The macroH2A nucleosome displays an alternate route in which pathways instead access the docking domain region via protein-DNA interactions. The H2A.Z system uniquely passes information along the H2B α2 helix of the opposing dimer. Pathways in H2A.Z also access the protein-DNA type route of macroH2A and the four helix bundle route of the canonical and L1-Mutant systems, but at a reduced frequency.

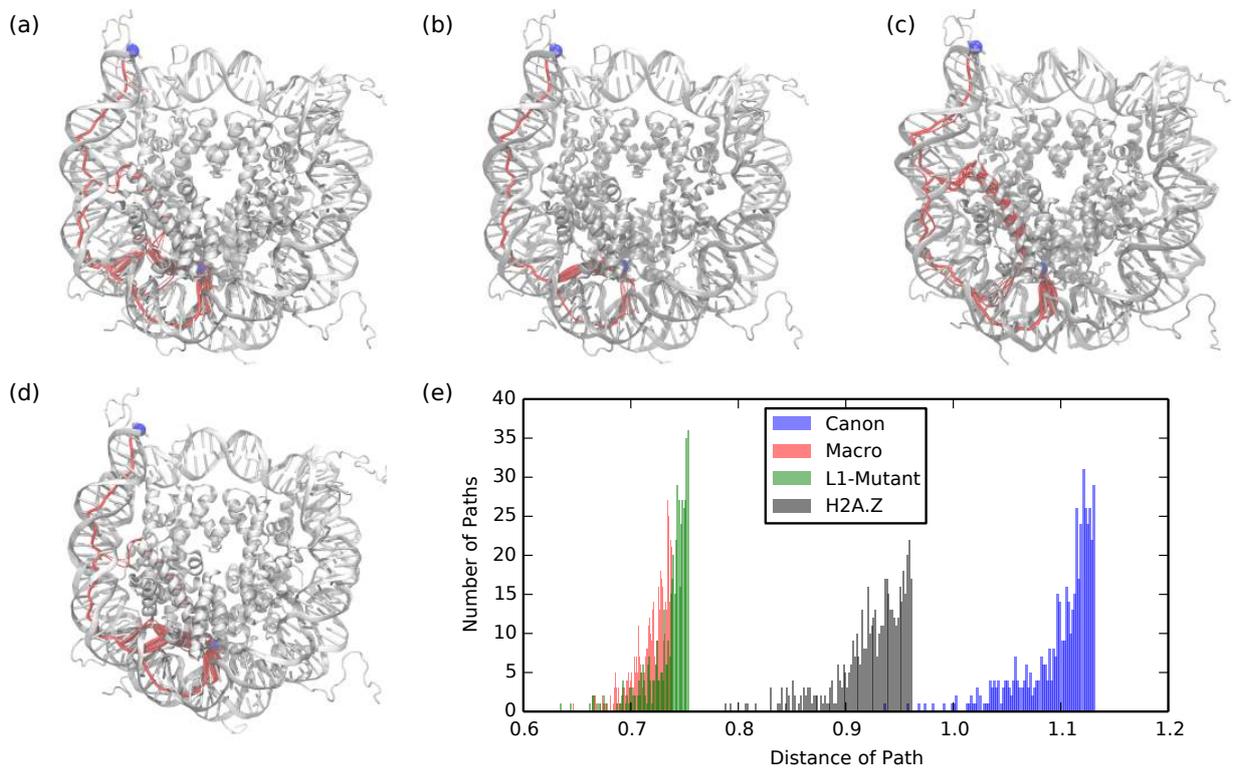

**Figure 6:** The 500 sub-optimal pathways between the L1 loops and symmetrically associated DNA entry for the (a) canonical NCP, (b) L1-Mutant, (c) macroH2A, and (d) H2A.Z projected on simulation snapshots. Also shown is the histogram of pathway distances (e). The L1 and DNA sites are represented as blue spheres, and the pathways are outlined in red with the wider pathways representing those of shorter "distance." Pathways in the variants are shorter, and thus stronger, than in the canonical NCP.

## PTM Targets are Located at Allosteric Hotspots

Beyond these specific pathways, dynamic networks exist throughout the NCP. To discern the importance of individual residues on these global networks, the edge-betweenness centrality of nucleosome residues was computed (Figure S8).[41] In the canonical NCP, a majority of the optimal pathways rely heavily on the DNA basepairs and neighboring histone tail lysine and arginine residues to propagate communication throughout the system. In H2A.Z, an increased number of shortest pathways access the L1 loops and the H4 α2 helix, but there remains a heavy reliance on the DNA and histone tails. In the L1-Mutant and macroH2A nucleosomes, dynamic traffic to the four-helix bundle increases. Furthermore, pathways in these systems access L1 residues more frequently than any other region.

Interestingly, we find that residues with the highest edge-betweenness scores are more likely to be the sites of post-translational modifications (PTMs). Based on the distribution of centrality scores, we classify residues in the upper tenth percentile as "hotspots" for communication (see supplemental). A comparison of known PTM sites with these allosteric hotspots indicates that PTMs are enriched at these locations, with an enrichment factor of 254% (p-value of 0.0155). When we compare our "hot spot analysis" with known PTM sites,[51–53] we observe a significant population of PTM targets (Figure 7). While PTMs in the histone core are identified more frequently than those in the tails, the most significant subset contains PTMs that have been implicated in affecting mononucleosome stabilities ("monoNCP PTMs").[51] In relation to the types of PTMs, methylation sites are linked with allosteric hotspots more frequently than phosphorylations or acetylations, likely due to their presence at DNA entry/exit sites and between turns of superhelical DNA.

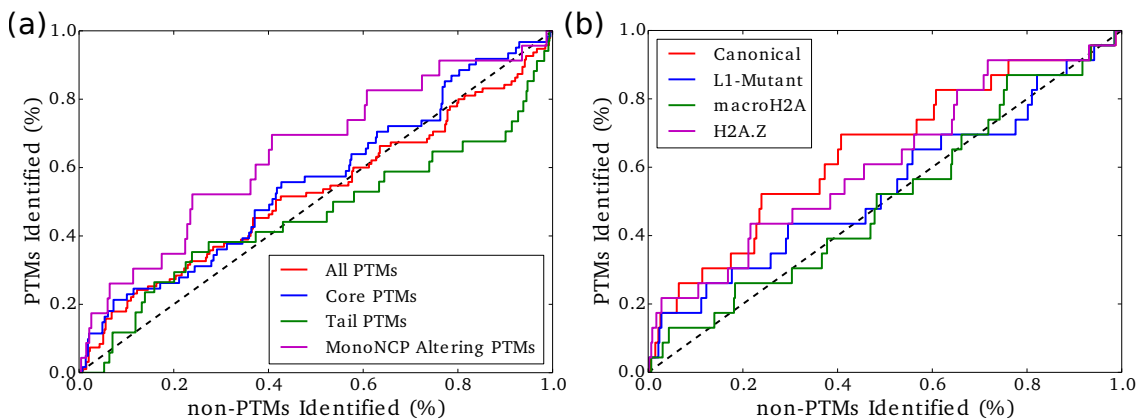

**Figure 7:** (a) Receiver Operator Curves (ROC) for subsets of PTM sites in the canonical nucleosome as identified by edge-betweenness centrality ranking. The largest enrichment can be seen in the PTM subset of monoNCP altering PTMs. The core PTMs are also more frequently identified than the tails. (b) ROC for monoNCP altering PTMs across the variant systems. The canonical and H2A.Z systems are shown to depend greater on monoNCP altering PTMs than the L1-Mutant and macroH2A systems for distributing dynamic information.

An overall correlation between allosteric hotspots and PTM locations is maintained in the nucleosome variants, however the specific details differ between the systems (Table S2). For example, all four systems show the importance of PTM sites in the H3 histone near the DNA extremities, while histone H4 monoNCP PTM sites in the four-helix bundle are accessed more frequently in variant networks. In general, the canonical system displays the greatest reliance on monoNCP PTMs, then the H2A.Z nucleosome, and finally the L1-Mutant and macroH2A systems, respectively.

## Structural Stability in Variant Nucleosomes

To quantify global NCP dynamics, a full correlation analysis (FCA) was performed on the $C_\alpha$ atoms of the histone core α-helices.[54] Two of the dominant motions identified in this analysis corresponded to the nucleosome opening motions described by Böhm *et al.*[24] Projections into the phase-space described by these and other FCA modes showed that all nucleosome systems had similar global dynamics on the hundreds of nanoseconds timescale (see Supplemental Figure S6). However, given that the dynamics of nucleosome opening likely occur on the millisecond timescale, our simulations are far too short to effectively explore the effects of histone variants on large-scale NCP dynamics.

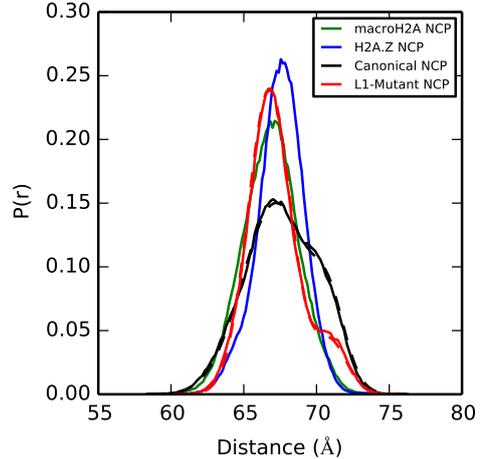

**Figure 8:** Distance populations for DNA end-to-end spread. The canonical system exists in two states: one centered around 67 Å ("compact") and one centered around 71 Å ("open"). While the L1-Mutant samples both states, the amount of time spent in the open state is drastically reduced. The macroH2A and H2A.Z nucleosomes exist only in the compact state. Fits are represented in dotted lines.

In contrast to the nucleosome opening motions, the DNA end-to-end separation distance does depend on the identity of the H2A histone (Figure 8). The sampling in the canonical system can be divided into two states: the prominent "compact" state centered around 67 Å and the "open" state centered at 71 Å. The L1-mutant sampled both the open and closed states, however the percentage of time spent in the open state was reduced from 14% to 11% of the simulation. Both the macroH2A and H2A.Z systems only sampled the closed state.

Results of an MM/GBSA analysis indicate that the overall DNA binding energetics are also altered by H2A variants. The DNA binding affinities to the L1-Mutant and macroH2A octamers were $31.0 \pm 9.1$ kcal/mol and $5.7 \pm 9.7$ more favorable than binding to the canonical NCP (Table 3). There are two primary contributors to this shift: direct interactions with the L1 loops ($\Delta\Delta G_{L1}$) and the changes in the DNA configuration ($\Delta\Delta G_{DNA}$). In the L1-mutant and macroH2A systems, $\Delta\Delta G_{L1}$ was largely a result of removing the negatively charged Glu[41] from the canonical loop and introduction of the Lys[40]-DNA interaction, which combine for an increase in binding free energy on the order of 10 kcal/mol. The reorientation of the dimers in the macroH2A-like systems also influences a favorable shift in DNA conformation relative to the canonical system ($\Delta\Delta G_{DNA}$ = -9.3 $\pm$ 5.6 kcal/mol and -17.0 $\pm$ 1.8 kcal/mol for the L1-Mutant and macroH2A nucleosomes, respectively). However, in macroH2A a number of small shifts in the remainder of the NCP oppose binding and therefore make it more comparable to the canonical system, which does not occur in the L1-mutant NCP. The H2A.Z nucleosome does not exhibit the same favorability for DNA binding when compared to the canonical octamer, but instead shows a disfavoring shift of 5.6 ± 8.9 kcal/mol. The removal of the negative charge on Glu[41] creates a favorable shift of 5.8 ± 0.1 kcal/mol in $\Delta\Delta G_{L1}$, but this is balanced by the nearly identical free energy penalty in the DNA rearrangement term $\Delta\Delta G_{DNA}$.

Table 3: MM/GBSA calculated binding energies for DNA binding affinity to the histone core in each of the NCP systems. $\Delta\Delta G$'s are referenced against the canonical NCP. The L1 loop sequence and DNA conformations of the variant structures contribute significantly toward favorable binding of DNA in the macroH2A-like systems, relative the canonical NCP. All values are reported in kcal/mol.

| System | $\Delta G_{binding}$ | $\Delta\Delta G_{binding}$ | $\Delta\Delta G_{L1}$ | $\Delta\Delta G_{DNA}$ | $\Delta G_{assembly}$ | $\Delta\Delta G_{assembly}$ |
|---|---|---|---|---|---|---|
| Canonical NCP | -428.6 ± 5.6 | —— | —— | —— | -618.6 ± 5.8 | —— |
| L1 Mutant | -459.6 ± 7.2 | -31.0 ± 9.1 | -8.7 ± 0.1 | -9.3 ± 5.6 | -679.6 ± 7.5 | -60.0 ± 9.5 |
| macroH2A NCP | -434.5 ± 7.9 | -5.7 ± 9.7 | -10.7 ± 0.1 | -17.0 ± 1.8 | -637.1 ± 7.9 | -18.5 ± 9.8 |
| H2A.Z NCP | -423.0 ± 6.9 | 5.6 ± 8.9 | -5.8 ± 0.1 | 5.8 ± 5.4 | -616.3 ± 6.9 | 2.3 ± 9.0 |

Similarly, another MM/GBSA analysis revealed that macroH2A variants modify the energetics of complex assembly. The $\Delta G_{assembly}$ of the L1-Mutant and macroH2A nucleosomes were -60.0 ± 9.5 and -18.5 ± 9.8 kcal/mol more favorable than the canonical system. The favorability in the macroH2A-like systems is a result of favorable DNA binding coupled with stronger protein-protein interactions. The $\Delta G_{assembly}$ of H2A.Z was in agreement with that of the canonical nucleosome (Table 3).

# Discussion

The simulations and analysis presented here detail a series of mechanisms by which the histone variants macroH2A and H2A.Z influence the dynamics of the nucleosome core particle. The subtle structural rearrangements these variants cause leverage the tightly packed nature of the histone core to influence the global energetics and dynamics of the complex, thus influencing gene expression. Dynamic effects appear to be particularly important, as they allow for the propagation of information through allosteric networks that span large distances. Although our simulations are only able to probe the sub-µs timescale, the dynamic differences observed at the dimer-tetramer and DNA/histone interfaces will likely be amplified on the ms timescale and result in these variants having altered nucleosome opening and DNA breathing motions.

These results also offer new insights into biochemical experiments that probed the mechanism of macroH2A. For example, Nusinow *et al.* showed that the L1-mutant is enriched in the inactive female X chromosome at nearly the same rate as the complete histone-domain of macroH2A.[11] Point mutations demonstrated that enrichment was significantly increased by the two mutations that introduce additional bulk into the L1 loops, N38H and E41Y, whereas it was decreased by the Y39P mutation, which decreases the size of the L1 loop. Based on our results, we believe that larger sidechains may help encourage the α2 "bulging" motion observed in each of the variant simulations, and therefore make the NCP more variant-like.

In another set of experiments, Chakravarthy *et al.* demonstrated that mutations to the L1 loops modulate the salt-dependent stability of the histone octamer.[21] They showed that in both the L1-mutant and macroH2A-containing system, the histone octamer is stable down to 0.5 M NaCl, whereas the canonical and H2A.Z-containing structures dissociate into dimer and tetramer constituents in solutions below 1.1 M. In agreement with this, we observe a significantly more favorable interaction between the H2A L1 loops in the variant structures than in the canonical structure. Since the L1-L1 interface is the only location of dimer-dimer interaction, stability in this region translates to octamer stability.

The mechanisms of H2A.Z remain more elusive. Stability studies have been non-conclusive as some indicate that H2A.Z enhances stability,[55] while others suggest that it destabilizes the nucleosome. Our simulations show H2A.Z nucleosome stabilities that are in agreement with the canonical system, despite their differing dynamics. These systems were constructed with identical sequences, except for H2A composition. Therefore, our findings support a mechanism which suggests that H2A.Z by itself has little-to-no effect on NCP stability. Instead, H2A.Z presence may be combined with other factors - such as PTMs or H3 variant presence - in order to alter particle stability.[22,56] Furthermore, the altered dynamics and locations of allosteric networks and hotspots between H2A.Z and canonical nucleosomes may result in different responses to these chromatin remodeling factors. The dimer reordering may also act to recruit transcriptional machinery to chromatin possessing large populations of H2A.Z, such as transcriptional starting sites.

Finally, we present a comprehensive analysis of the dynamic networks in the nucleosome. We observe that these networks are strongly affected by the dynamics of the L1 loops which are allosterically linked to a wide number of important regions in the nucleosome core. Using only small changes in their structure, variants are able to modify these networks to affect the function of the NCP. We hypothesize that this is a general mechanism that other chromatin remodeling factors may also utilize. For example, the finding that PTMs are enriched at residues with increased allosteric activity suggests that these perturbations may take advantage of dynamic networks to amplify their effects on chromatin and influence global NCP dynamics. In addition, by altering these networks, variants may be able to tune the responses of nucleosomes to specific PTMs. Future work to study the disparate effects of chromatin remodeling factors on dynamics in the nucleosome is required to fully understand the mechanisms of *in vivo* gene expression and regulation.

# Supplementary Information

Complete methods; Tables S1-S2; Figures S1-S23.

## Acknowledgement

The authors thank S. Chakravarthy for valuable discussions concerning the work presented here. Research reported in this publication was supported by the National Institute of General Medical Sciences of the National Institutes of Health [grant no. R15GM114758]. The content is solely the responsibility of the authors and does not necessarily represent the official views of the National Institutes of Health. This work used the Extreme Science and Engineering Discovery Environment (XSEDE), which is supported by National Science Foundation grant number ACI-1053575. In addition, this research used resources of the National Energy Research Scientific Computing Center, which is supported by the Office of Science of the U.S. Department of Energy under Contract No. DE-AC02-05CH11231.

# Supporting information for:

# Effects of macroH2A and H2A.Z on nucleosome structure and dynamics as elucidated by molecular dynamics simulations

Samuel Bowerman and Jeff Wereszczynski*

*Department of Physics and Center for Molecular Study of Condensed Soft Matter, Illinois Institute of Technology, 3440 S Dearborn St., Chicago, IL 60616, USA*

E-mail: jwereszc@iit.edu

## Methods

### System Construction and Simulation Details

The canonical nucleosome was initiated from the crystal structure of Daveys *et al.* (PDB ID: 1KX5).[S1] The crystallographic $Mn^{2+}$ were replaced by physiological $Mg^{2+}$. Additional $Mg^{2+}$ ions were added to fill symmetrically suggested voids. The crystallographic waters were also maintained. The L1-Mutant structure was then created from the canonical one by mutating the [38]NYAE[41] H2A L1 loops to [38]HPKY[41] sequence of macroH2A. The mutation was done using VMD. The macroH2A system was initialized from the crystal structure solved by Chakravarthy *et al.* (PDB ID: 1U35).[S2] The missing tail segments were constructed using the canonical structure as a reference. Strong similarities in DNA arrangement - particularly at DNA-protein binding sites (Figure S1) - allowed for the 146 basepairs of DNA from the crystal structure to be replaced by the 147 bp (plus $Mg^{2+}$) of the 1KX5 structure, and H3 residues were mutated to match the sequence of the 1KX5 system. These actions were taken to ensure that differences between the systems were attributable only to H2A sequence divergence. The H2A.Z system was constructed analogously, using the crystal structure of

---

*To whom correspondence should be addressed

Suto *et al.* (PDB ID: 1F66).[S3] Therefore, each system was composed of 147 palindromic basepairs of α-satellite DNA wrapped around a histone core of *Xenopus laevis* H3, H4, and H2B with human H2A histones and variants. Histidine states were assigned using PROPKA and the interactive H-Bond Optimizer of the Desmond-Schrödinger package.[S4]

Each system was simulated three times. Each simulation underwent 10,000 steps of geometric minimization (5,000 steps with protein heavy atoms harmonically restrained by a force constant of 10 kcal/mol/Å$^2$ and 5,000 steps without restraints). Heating was done by gradually raising the temperature from 10 to 300 K over 6 ps of simulation in the NVT ensemble. During heating, protein heavy atoms were harmonically restrained with a force constant of 10 kcal/mol/Å$^2$. The restraints were then gradually released over 600ps in the NPT ensemble.[S5] Each simulation was then conducted for an additional 250ns in the NPT ensemble using a Langevin piston with a 100 fs period and collision frequency of 3 ps$^{-1}$. The SHAKE algorithm was used to allow for a 2 fs timestep, and long-range electrostatics were calculated using the particle mesh Ewald method.[S6,S7] Short range interactions were calculated with a 10 Å cutoff, where a switching function was applied at 8 Å. It was observed that ∼50ns was required for system equilibration, and so ∼200ns of production data was obtained from each simulation (600ns per system).

## Allosteric Calculations

Residue correlations were calculated using the "largest linear mutual information" method.[S8,S9] In this method, the linear mutual information is calculated between all heavy atoms in the system. The residue-wise mutual information values were converted to a Pearson correlation coefficient-like value by $r_{i,j} = [1 - e^{(-2I_{i,j}/3)}]^{1/2}$, where $I_{i,j}$ is the largest linear mutual information between any two atoms of the residues $i$ and $j$.

Contact maps were produced in-house using the MDAnalysis package.[S10] Two residues were considered to be in contact if their Cα (protein) or C1' (nucleic) atoms were within 10 Å for 70% of the configurations. Using the predefined correlation matrices and this contact map, a NetworkX edgelist was formed.[S11] The length of each edge was defined by $D_{i,j} = -log(r_{i,j})$, where r$_{i,j}$ is the correlation value between residues $i$ and $j$. The optimal paths were calculated using the NetworkX implementation of Dijkstra's algorithm, and the suboptimal paths were determined using Yen's K-Shortest Paths algorithm.

The Kullbach-Leibler Divergence of dihedral angles were calculated using the method of McClendon.[S12] In this method, each simulation was separated into three blocks (9 blocks per system) with 31,666 configurations (roughly 63.3 ns) per block. Histogram widths were 15 degrees. The Kullbach-Leibler Divergence values were calculated for each variant by using the populations of the canonical system as a reference set. Bootstrapping techniques were employed to calculate the self-divergence of the canonical system. For any residue in a variant system whose divergence value was below this self-divergence threshold, the KL-Divergence value for that residue was set to 0.

## Edge-betweenness Centrality

The importance of a node in a communication network can be defined by its edge-betweenness centrality.[S13] In this method, the "shortest" correlation pathway between all residue-pairs is calculated. A residue's edge-betweenness centrality is then defined as the number of shortest paths in which the residue appears:

$$C(i) = \frac{1}{N} \sum_{s \neq t \neq i} x_i(s, t) \qquad (1)$$

where $N$ is the total number of paths and $x_i(s,t)$ is either 0 (residue $i$ does not exist in path between residues $s$ and $t$) or 1 (residue $i$ does exist in said path). For visualization purposes, centrality values are normalized such that the minimum centrality is 0 and the maximum centrality is 1, according to the formula:

$$C_{norm}(i) = \frac{C(i) - C_{min}}{C_{max} - C_{min}} \qquad (2)$$

where $C_{max}$ and $C_{min}$ are the maximum and minimum centrality values in the network.

From the plot of normalized centrality value vs percentile (Figure S9), we observe that the difference in centrality between percentile increments is not constant but is large at the upper and lower quartiles and steady in the interquartile region. The inflection point of the upper quartile exists near the tenth percentile, so we have chosen this location as our cut-off for defining allosteric "hotspots."

We were able to identify several well-known post-translational modifications (PTMs) in the canonical nucleosome by our betweenness centrality measurement.[S14–S16] In the canonical system, we observe 6 of 23 mononucleosome altering PTM sites (monoNCP PTMs) in the upper tenth percentile and 12 of 23 in the upper quartile. The significance of observing this subset of residues in each percentile was tested by calculating the pmf of a hypergeometric distribution,[S17]

$$\text{pmf}(x = k) = \frac{\binom{K}{k}\binom{N-K}{n-k}}{\binom{N}{n}} \qquad (3)$$

where $N$ is the total number of protein residues (487), $n$ is the percentile population size ($n=49$ for the upper tenth, and $n=122$ for the upper quartile), $K$ is the total number of monoNCP PTM residues (23), and $k$ is the number of observed monoNCP PTM sites ($k=6$ for the upper tenth, and $k=12$ for the upper quartile). Using these values, the upper tenth percentile observation has a p-value of 0.0155, and the upper quartile p-value is 0.00288. Therefore, the observation that monoNCP PTMs are located at allosteric hotspots is statistically significant.

Furthermore, we can quantify the presence of monoNCP PTMs at allosteric hotspots by calculating the enrichment factor (EF) of monoNCP PTMs over random selection,

$$EF = \frac{k}{K}\frac{N}{n} \qquad (4)$$

where the variables have the same meaning as for the hypergeometric distribution. We then calculate an EF of 2.54 for monoNCP PTM presence at allosteric hotspots. A plot of EF vs centrality percentile can be found in Figure S11.

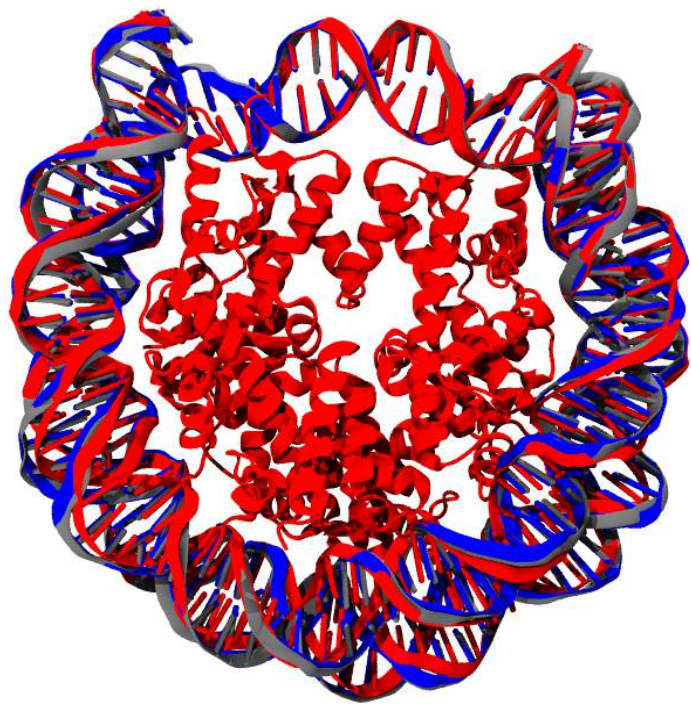

**Figure S1:** Comparison of crystallographic DNA arrangment in the canonical (blue), macroH2A (red), and H2A.Z (grey) nucleosomes. As expected, the DNA-histone binding sites show strong agreement in coordination between the three structures.

**Table S1:** Hydrogen bonding at key locations in the nucleosome. For the H-Bonds formed with the DNA, the occupancy of each bond is given. For the other interactions, the average number of hydrogen bonds between each group in a given frame is listed.

| System | H2A R$^{29}$-DNA | H2B S$^{33}$-DNA | L1-L1 | Dimer-tetramer |
|---|---|---|---|---|
| Canonical NCP | 63% | 40% | $0.5 \pm 0.1$ | $14.8 \pm 1.1$ |
| L1 Mutant | 70% | 60% | $0.4 \pm 0.1$ | $16.5 \pm 1.0$ |
| macroH2A NCP | 52% | 30% | $0.2 \pm 0.1$ | $14.7 \pm 0.8$ |
| H2A.Z NCP | 73% | 85% | $0.1 \pm 0.1$ | $14.4 \pm 1.1$ |

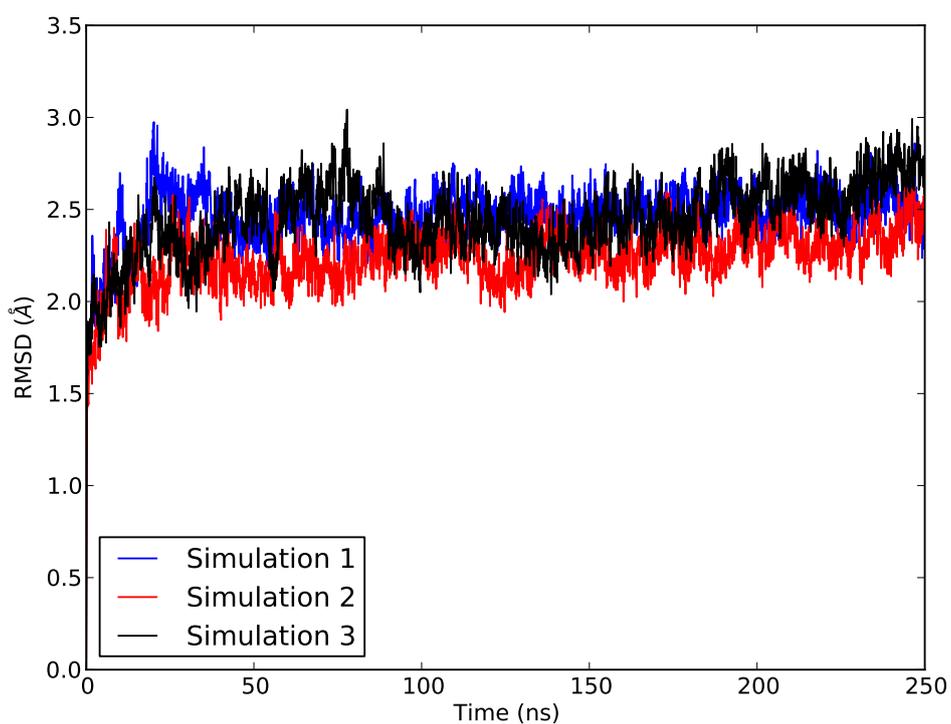

**Figure S2:** Backbone RMSD of three canonical simulations. The simulations were fit to the histone core backbone, and the RMSD calculations were done on the DNA and histone core backbone atoms, excluding tail residues.

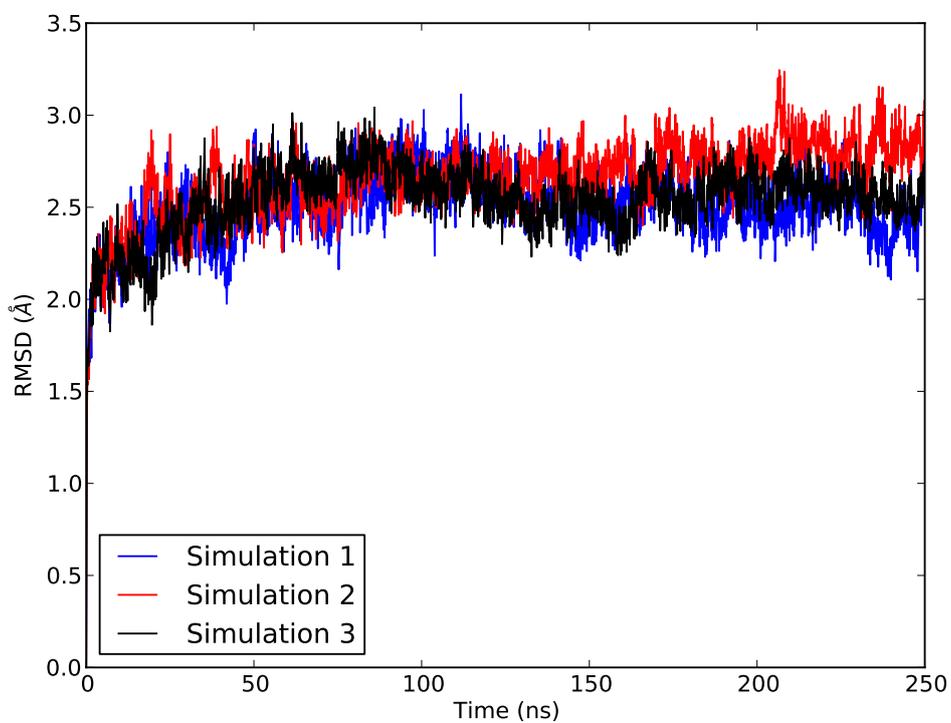

**Figure S3:** Backbone RMSD of three L1-Mutant simulations. The simulations were fit to the histone core backbone, and the RMSD calculations were done on the DNA and histone core backbone atoms, excluding tail residues.

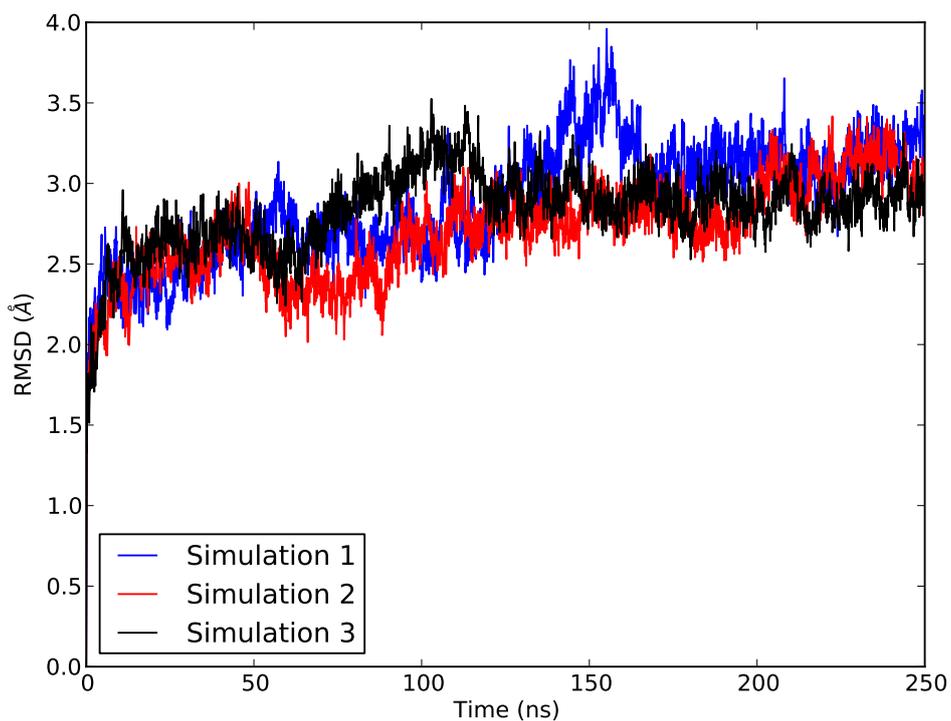

**Figure S4:** Backbone RMSD of three macroH2A simulations. The simulations were fit to the histone core backbone, and the RMSD calculations were done on the DNA and histone core backbone atoms, excluding tail residues.

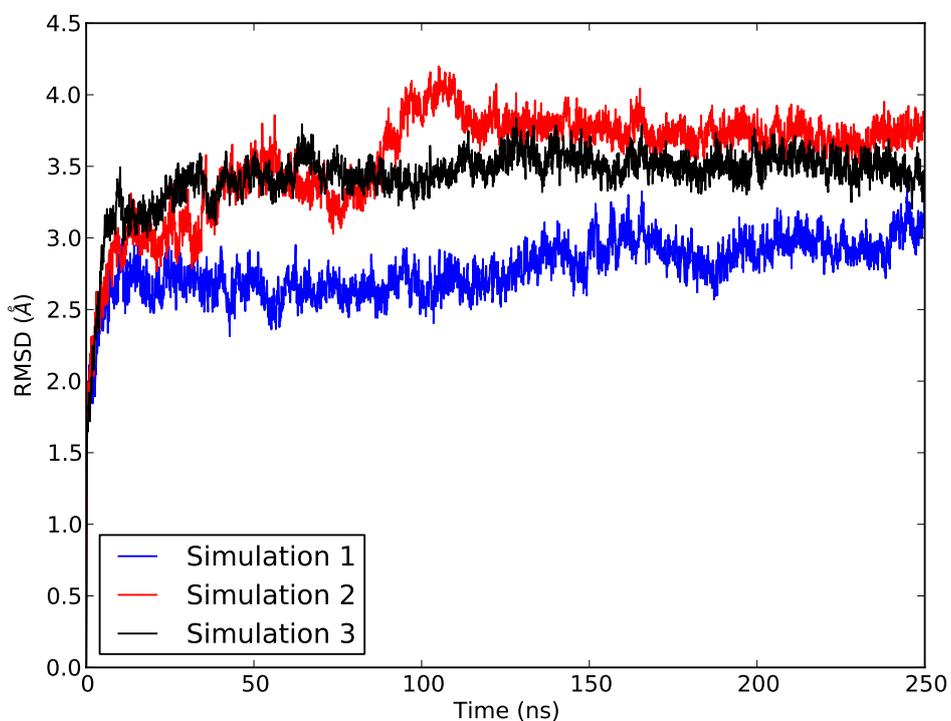

**Figure S5:** Backbone RMSD of three H2A.Z simulations. The simulations were fit to the histone core backbone, and the RMSD calculations were done on the DNA and histone core backbone atoms, excluding tail residues.

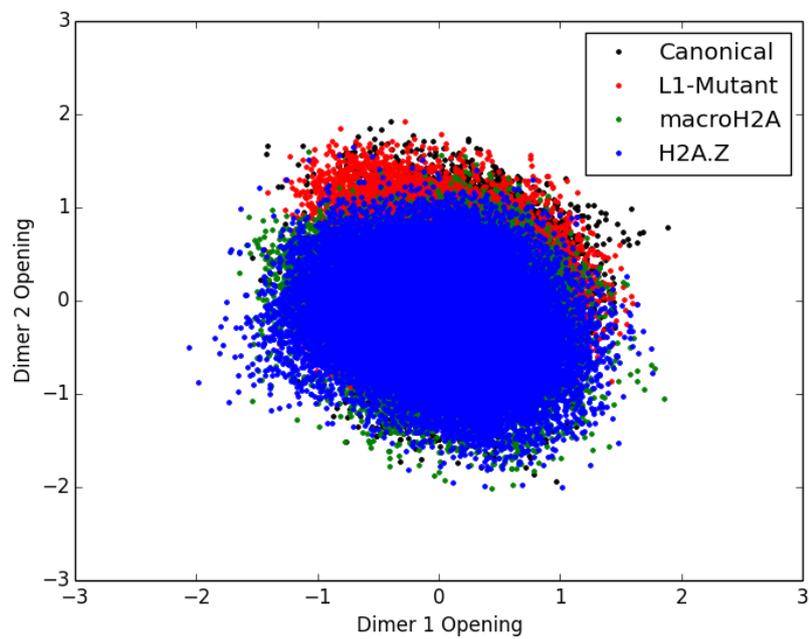

**Figure S6:** Dimer dissociation phase space for the nucleosome systems. All systems sample the same region of this space, which suggests that H2A composition has little effect on dimer dissociation in the hundreds of nanoseconds timescale. For reference, the dimer separation event occurs on the millisecond timescale.

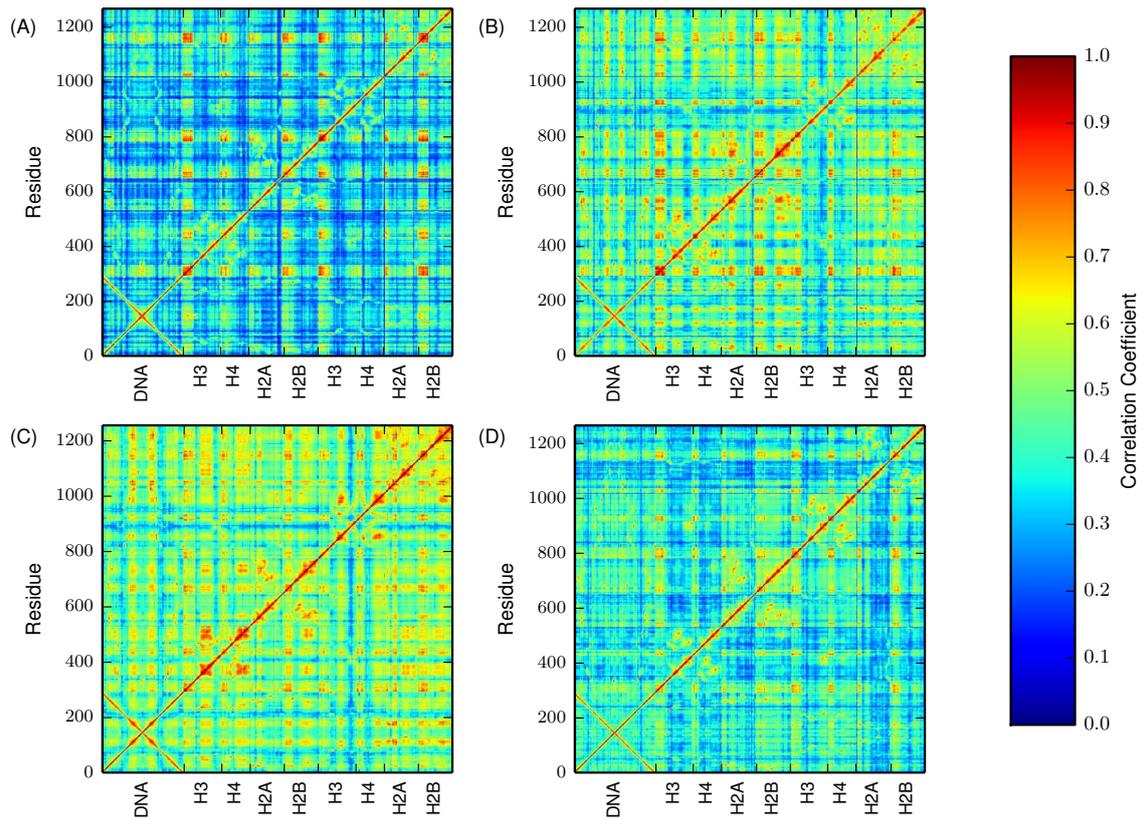

**Figure S7:** Pearsified Largest Linear Mutual Information matrices for (a) canonical, (b) L1-Mutant, (c) macroH2A, and (d) H2A.Z nucleosomes. The canonical NCP shows the weakest average correlation across the whole molecule, and the macroH2A variant shows the strongest. The L1-Mutant correlation strengths are similar to macroH2A, while the H2A.Z nucleosome shows correlations that are slightly above the levels of the canonical nucleosome. The steady increase in correlations within the variant systems is likely a result of favorable changes in interhistone interactions.

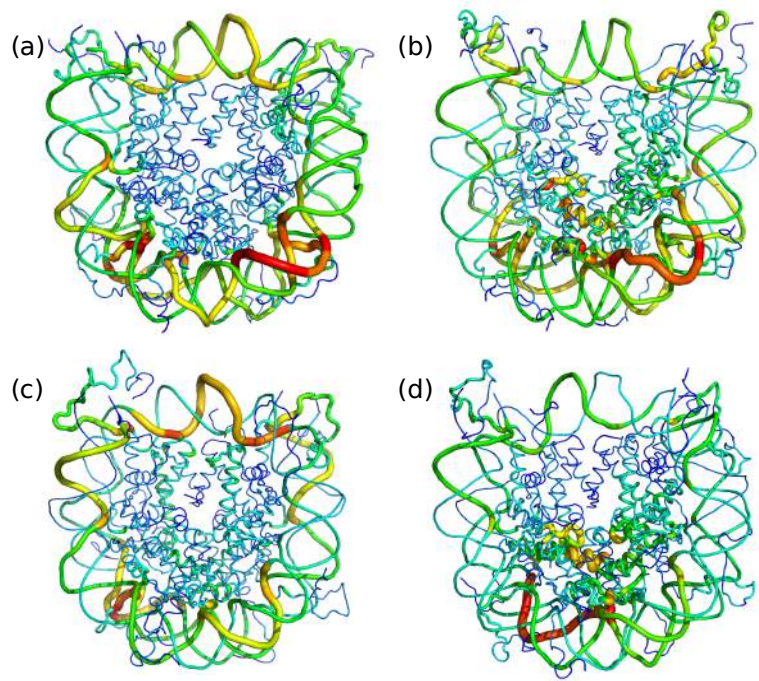

**Figure S8:** Edge-betweenness centrality for (a) canonical, (b) L1-Mutant, (c) H2A.Z, and (d) macroH2A nucleosomes. Brighter, wider regions represent locations that are accessed more frequently in the optimal communication pathways of each nucleosome system. The H2A L1-L1 interaction region in the L1-Mutant and macroH2A systems act as communication hubs for allosteric networks in the nucleosome, whereas the canonical and H2A.Z nucleosomes rely heavier on DNA to propagate information.

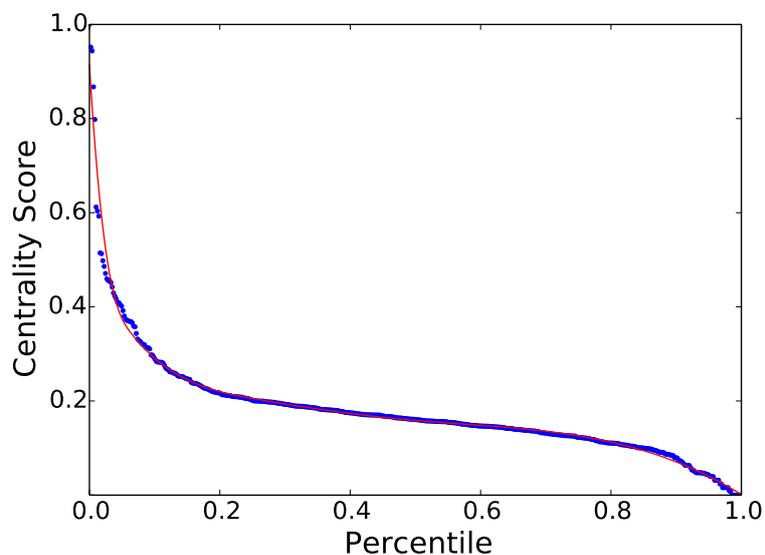

**Figure S9:** Centrality scores vs percentile ranking (blue dots) for the canonical nucleosome. The spline fit is represented in red. The drastic change in centrality score in the upper quartile indicates that residues rely heavily on several key residues for information propagation. The inflection point of this trend is located at the tenth percentile.

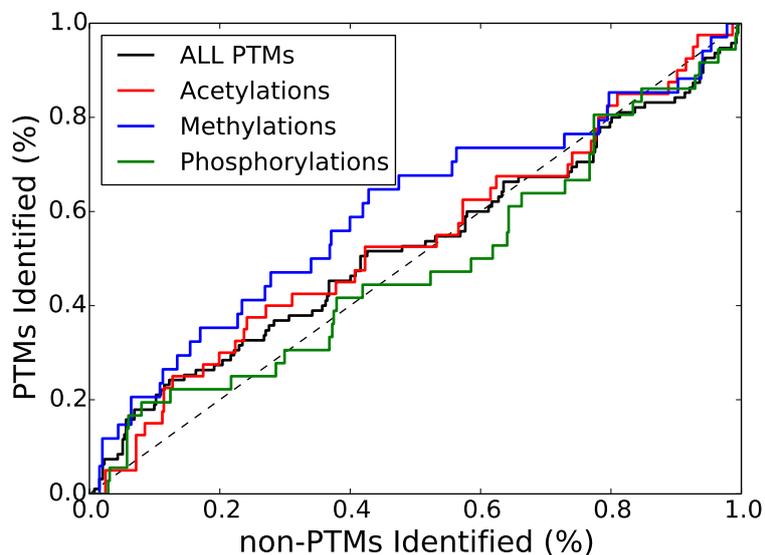

**Figure S10:** ROC plot for the canonical nucleosome with PTMs separated by modification type. Methylations exist most prevalently as allosteric hotspots, and acetylations are the least prevalent. The early enrichment of methylations is a result of their presence near DNA extremities and between superhelical DNA turns where pathways cross the symmetries.

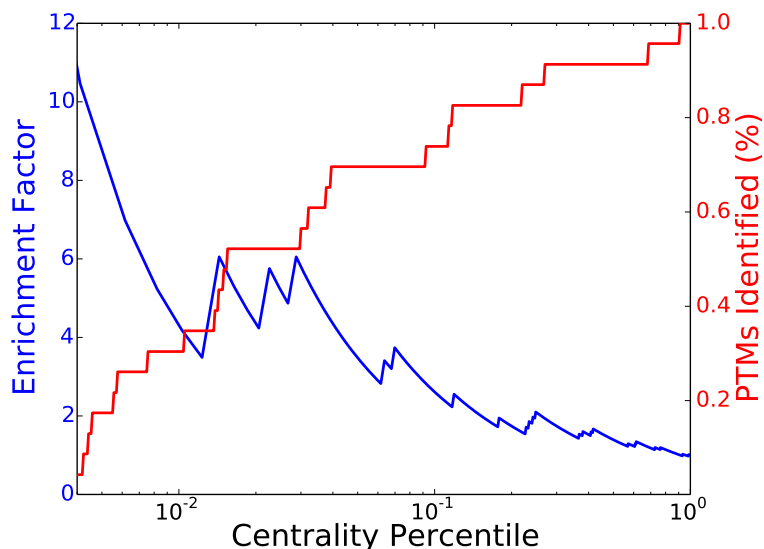

**Figure S11:** Enrichment Factor (blue) and monoNCP PTM identification percent (red) as functions of centrality percentile in the canonical nucleosome. We observe a strong degree of early enrichment for identifying monoNCP PTMs at allosteric hotspots. At the cutoff of the tenth percentile, we observe an EF of 2.54.

**Table S2:** Protein-normalized centrality values for the 23 monoNCP PTM targets. The percentile rank of each value is also listed. The residues in the upper quartile are listed in bold. Those in the upper tenth percentile are italicized. PTM sites that have significantly smaller centrality values than neighboring sequence residues in the upper quartile are labeled by an (*) and the value of the neighbor is reported.

|  | Canonical | | L1-Mutant | | macroH2A | | H2A.Z | |
| --- | --- | --- | --- | --- | --- | --- | --- | --- |
| PTM | Centrality | Percentile | Centrality | Percentile | Centrality | Percentile | Centrality | Percentile |
| H3 K4me3 | 0.05 | 6.6 | 0.07 | 6.3 | 0.05 | 6.2 | 0.06 | 6.7 |
| H3 K9ac | 0.13 | 27.0 | 0.19 | 22.6 | 0.14 | 24.8 | 0.16 | 27.5 |
| H3 K14ac | 0.20 | **75.6** | 0.30 | 51.0 | 0.23 | 53.4 | 0.26 | 61.2 |
| H3 K18ac | 0.27 | **87.9** | 0.39 | 73.9 | 0.30 | 69.8 | 0.34 | **78.4** |
| H3 K23ac | 0.34 | **92.8** | 0.51 | **87.3** | 0.38 | **82.0** | 0.44 | **88.9** |
| H3 K36me2,3 | 0.47 | *97.5* | 0.80 | *97.3* | 0.45 | **86.1** | 0.69 | *97.7* |
| H3 Y41ph | 0.46 | *96.9* | 1.00 | *100.0* | 0.67 | *96.9* | 0.84 | *99.8* |
| H3 R42me2a | 0.51 | *98.4* | 0.79 | *96.7* | 0.37 | **81.5** | 0.79 | *99.2* |
| H3 T45ph* | 0.29 | **75.0** | 0.38 | 71.1 | 0.22 | 52.4 | 0.28 | 67.0 |
| H3 K56ac | 0.17 | 58.9 | 0.18 | 20.1 | 0.22 | 52.4 | 0.19 | 35.6 |
| H3 S57ph | 0.15 | 39.0 | 0.16 | 17.9 | 0.22 | 52.4 | 0.16 | 28.6 |
| H3 K64ac | 0.15 | 42.7 | 0.26 | 38.0 | 0.27 | 63.6 | 0.21 | 46.1 |
| H3 K115ac | 0.15 | 38.4 | 0.13 | 11.7 | 0.06 | 8.5 | 0.24 | 53.9 |
| H3 T118ph | 0.37 | *93.4* | 0.43 | **81.9** | 0.14 | 25.8 | 0.63 | *96.5* |
| H3 K122ac | 0.23 | **81.9** | 0.29 | 47.4 | 0.13 | 23.9 | 0.39 | **82.7** |
| H4 K16ac | 0.21 | **77.0** | 0.38 | 70.0 | 0.16 | 33.9 | 0.18 | 34.2 |
| H4 S47ph* | 0.25 | **85.4** | 0.28 | 43.9 | 0.19 | 44.3 | 0.34 | **78.2** |
| H4 K77ac | 0.18 | 61.8 | 0.29 | 45.2 | 0.15 | 28.3 | 0.25 | 58.0 |
| H4 K79ac* | 0.31 | *91.4* | 0.34 | 63.8 | 0.56 | *95.0* | 0.35 | **79.8** |
| H4 K91ac | 0.12 | 23.4 | 0.38 | 70.0 | 0.57 | *95.4* | 0.29 | 68.9 |
| H4 R92me | 0.21 | **76.4** | 0.53 | **88.5** | 0.81 | *99.4* | 0.33 | **77.6** |
| H2B K30ar | 0.94 | *99.6* | 0.81 | *97.7* | 0.26 | 62.4 | 0.72 | *98.8* |
| H2B K123ub1 | 0.00 | 0.0 | 0.00 | 0.0 | 0.00 | 0.0 | 0.00 | 0.0 |

S13

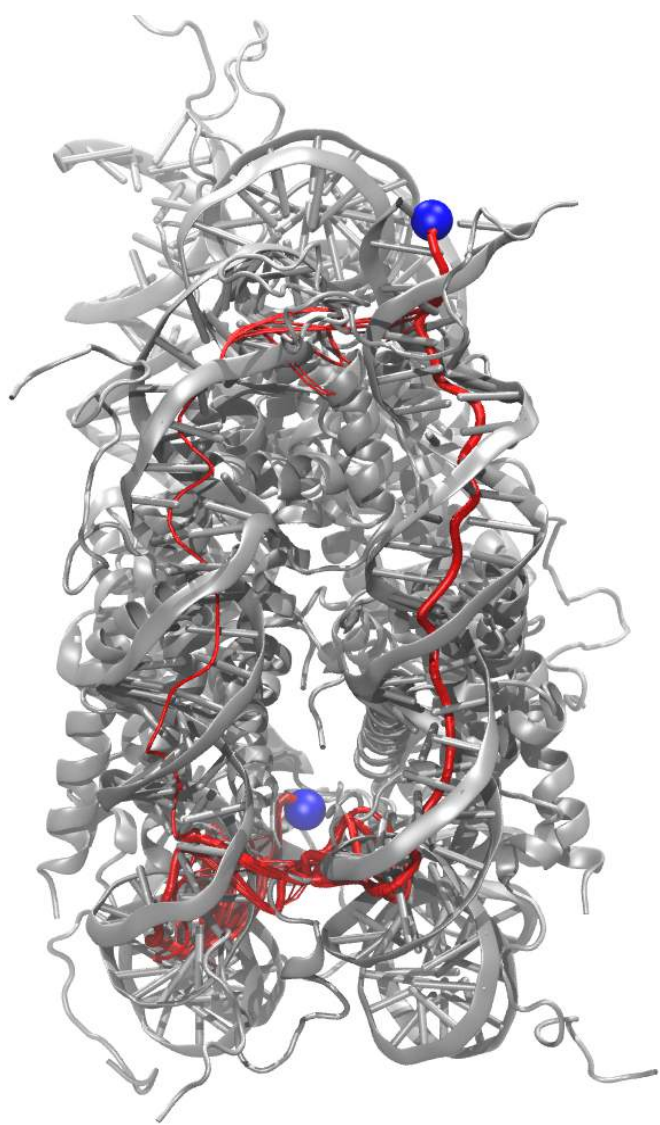

**Figure S12:** Sub-optimal pathways for L1 loop to DNA exit in the canonical nucleosome.

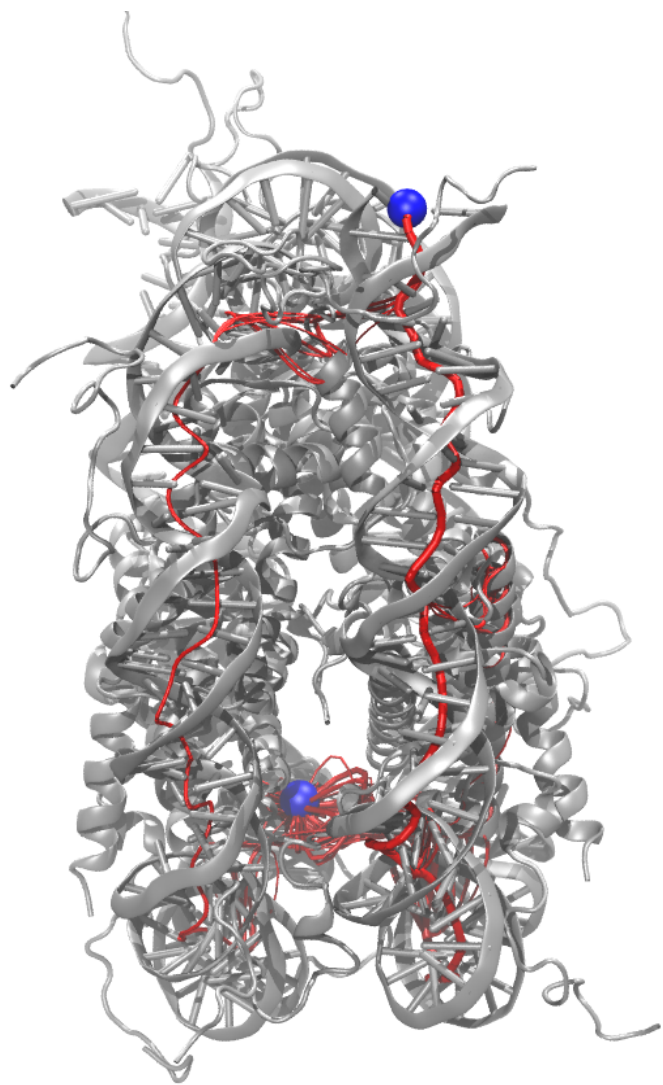

**Figure S13:** Sub-optimal pathways for L1 loop to DNA exit in the L1-Mutant nucleosome.

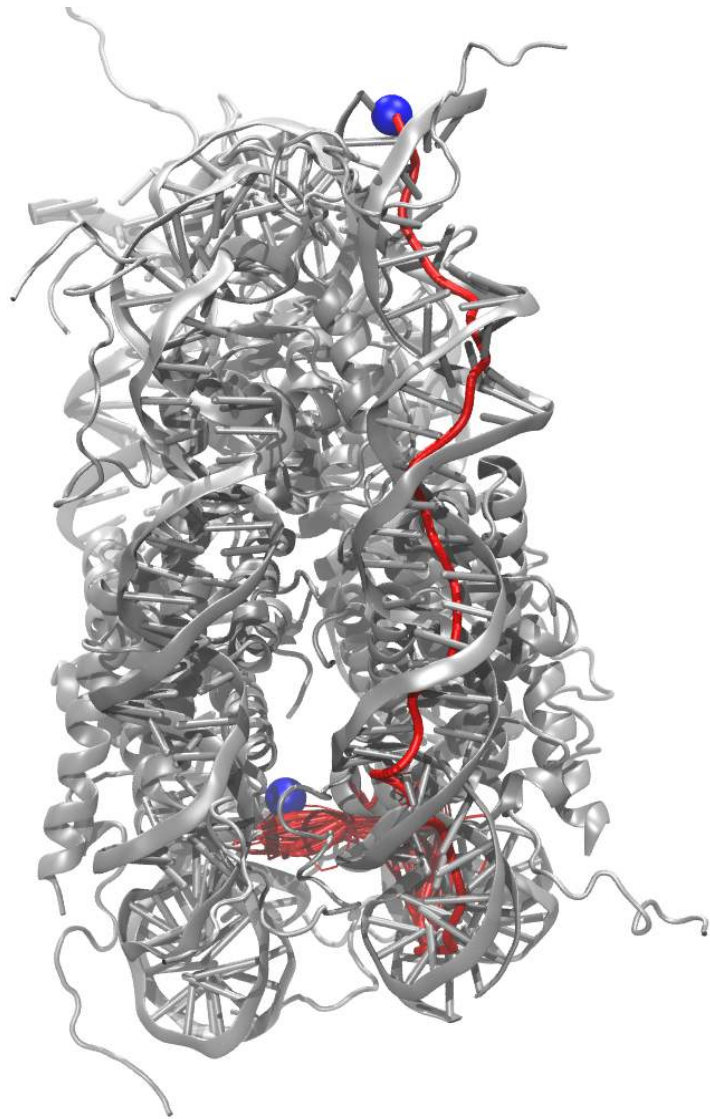

**Figure S14:** Sub-optimal pathways for L1 loop to DNA exit in the macroH2A nucleosome.

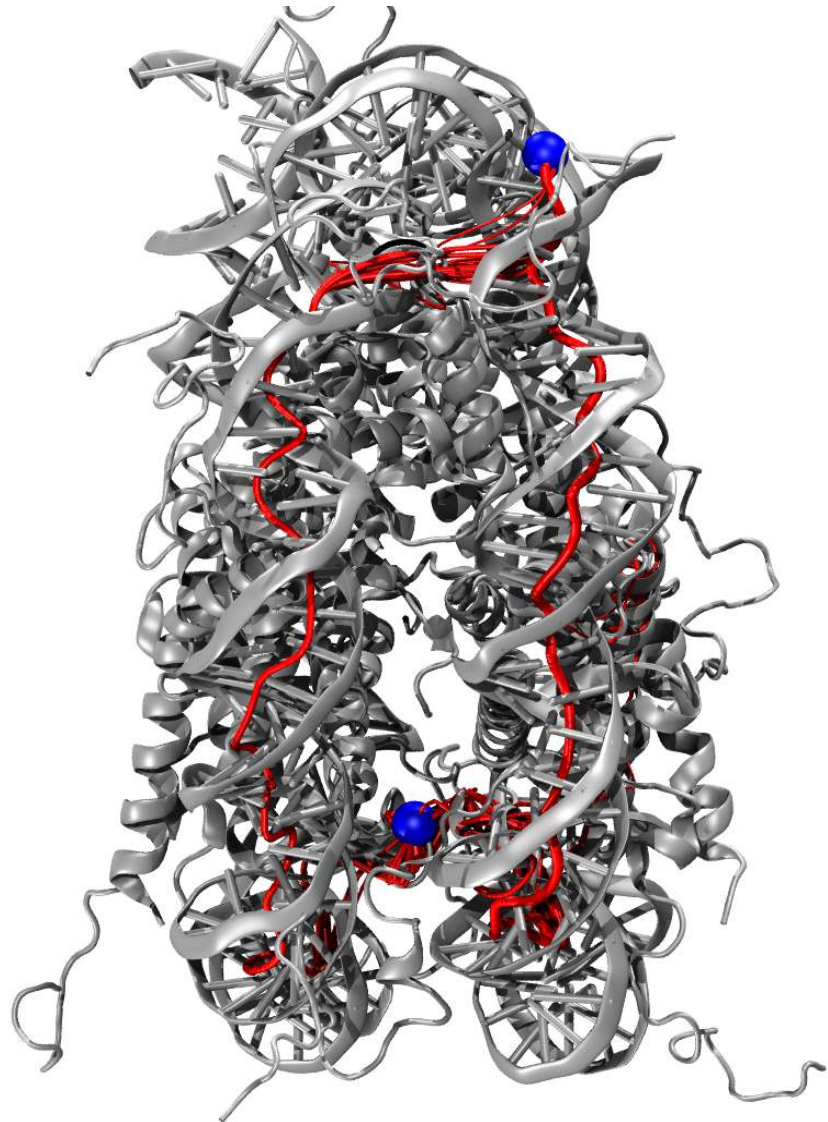

**Figure S15:** Sub-optimal pathways for L1 loop to DNA exit in the H2A.Z nucleosome.

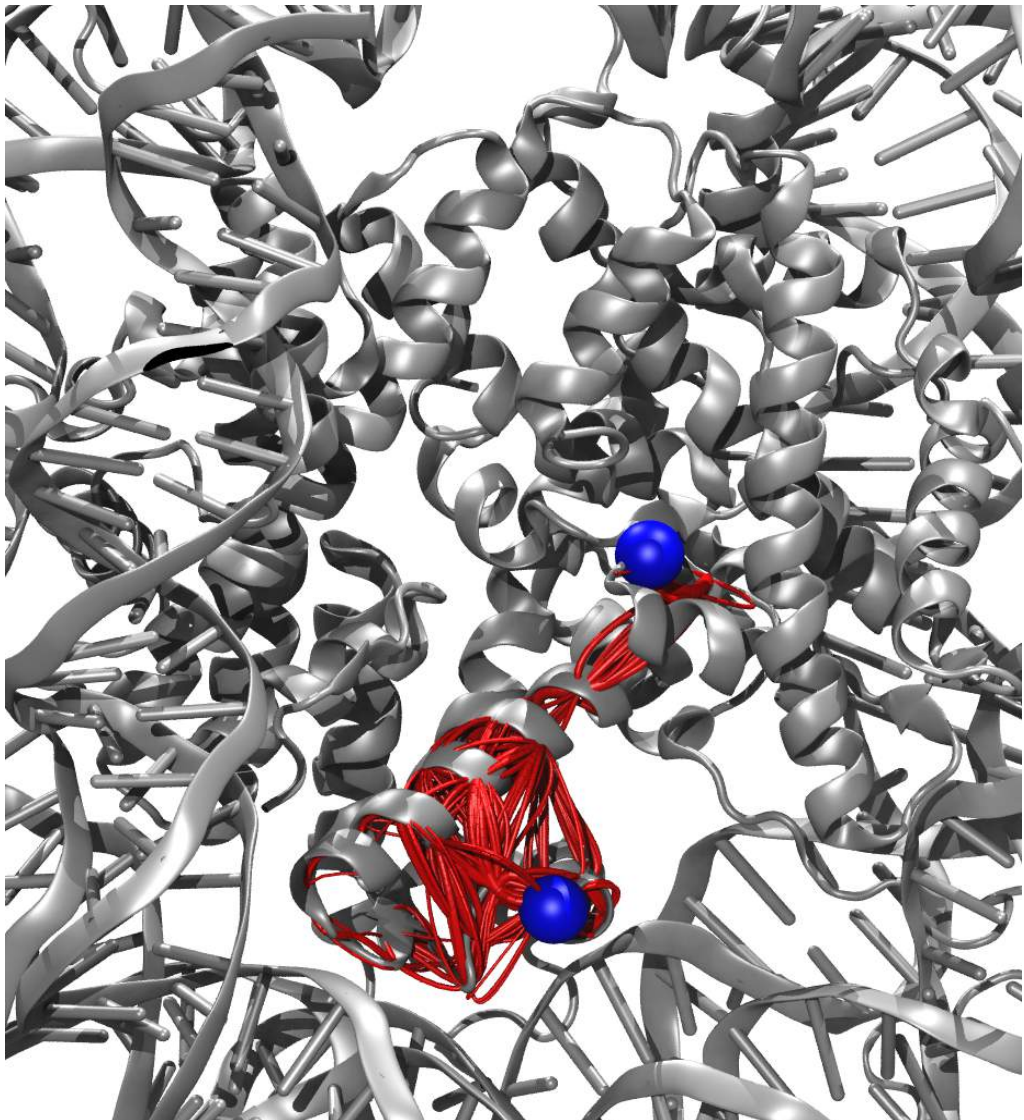

**Figure S16:** Sub-optimal pathways for L1 loop to the associated docking domain in the canonical nucleosome. The opposing dimer has been removed to improve visualization clarity.

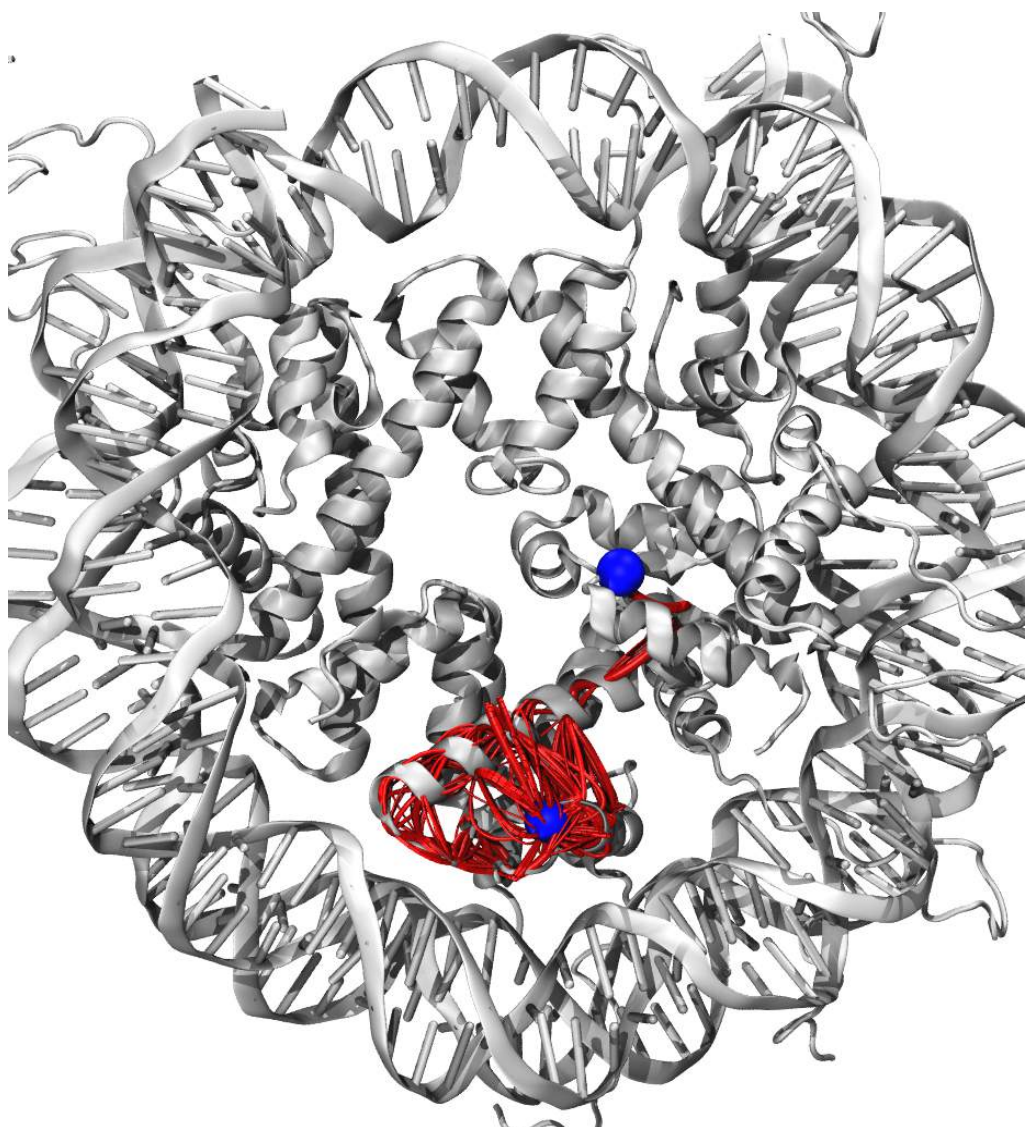

**Figure S17:** Sub-optimal pathways for L1 loop to the associated docking domain in the L1-Mutant nucleosome. The opposing dimer has been removed to improve visualization clarity.

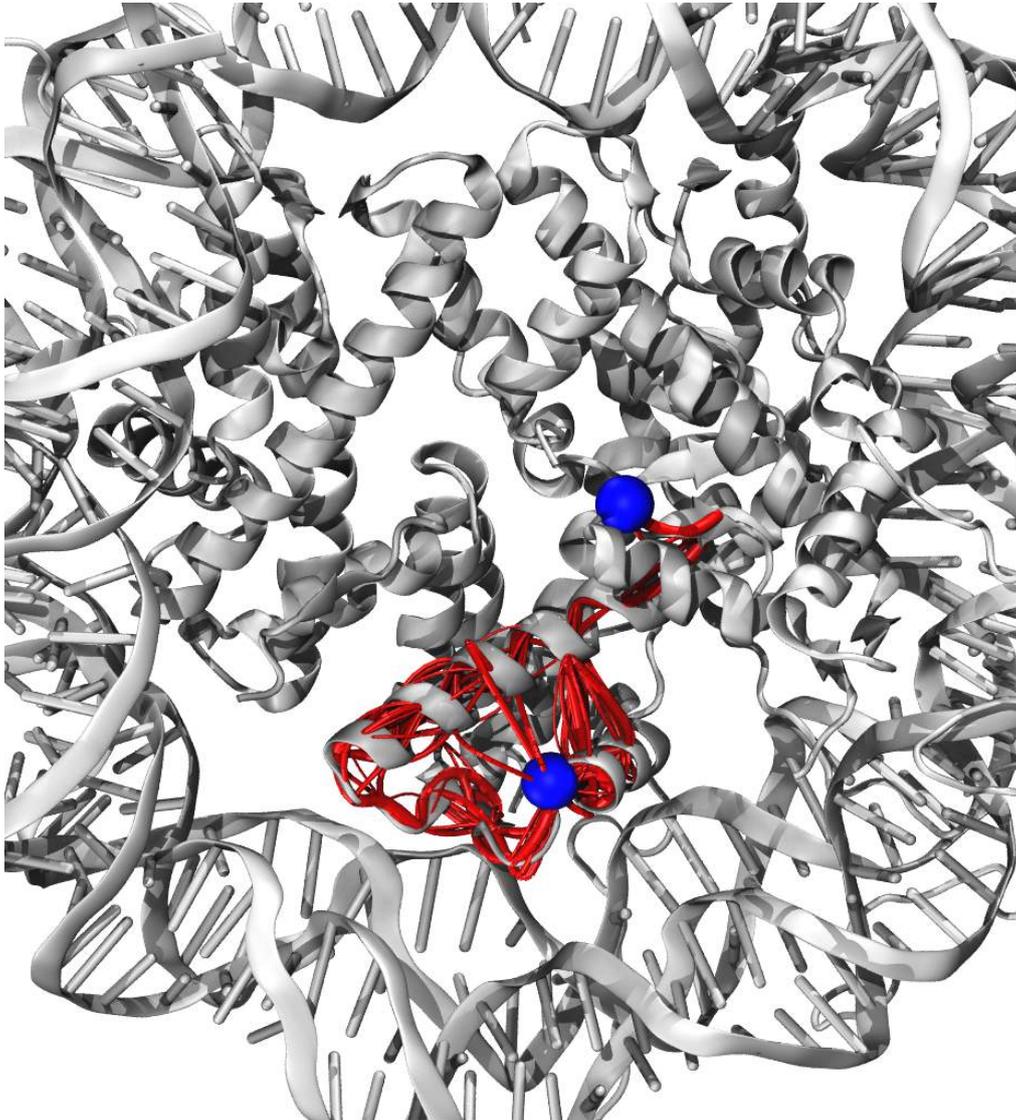

**Figure S18:** Sub-optimal pathways for L1 loop to the associated docking domain in the macroH2A nucleosome. The opposing dimer has been removed to improve visualization clarity

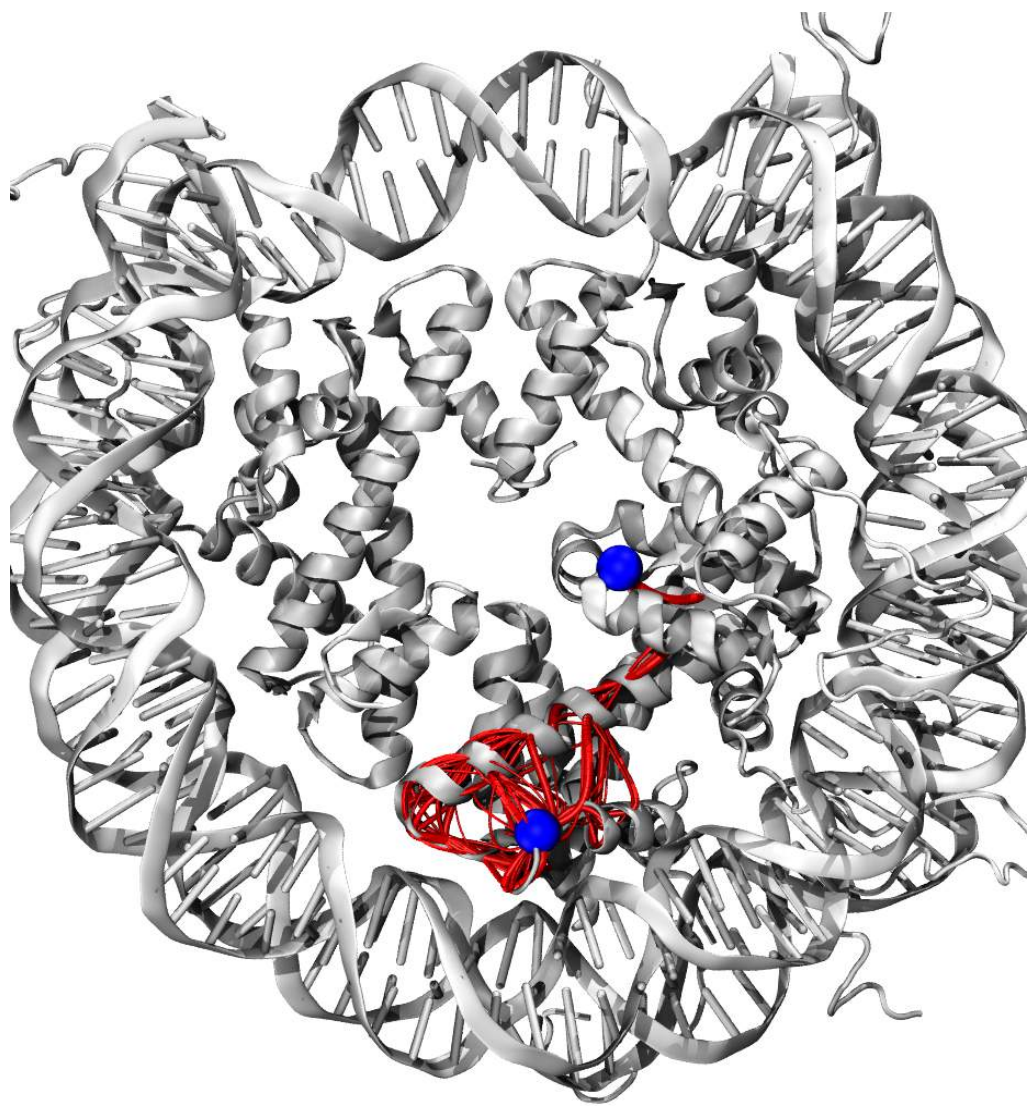

**Figure S19:** Sub-optimal pathways for L1 loop to the associated docking domain in the H2A.Z nucleosome. The opposing dimer has been removed to improve visualization clarity.

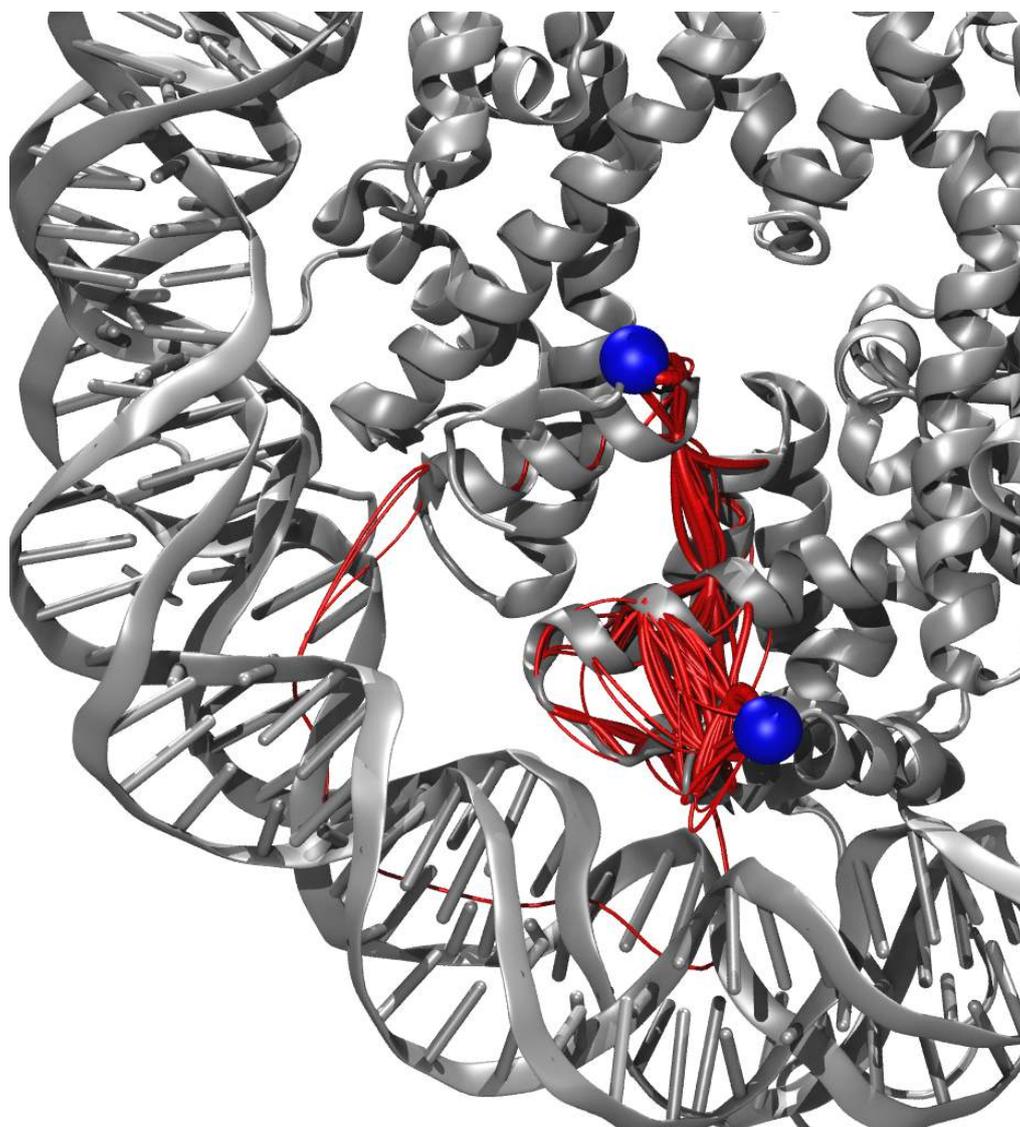

**Figure S20:** Sub-optimal pathways for L1 loop to non-associated docking domain in the canonical nucleosome. The opposing dimer has been removed to improve visualization clarity.

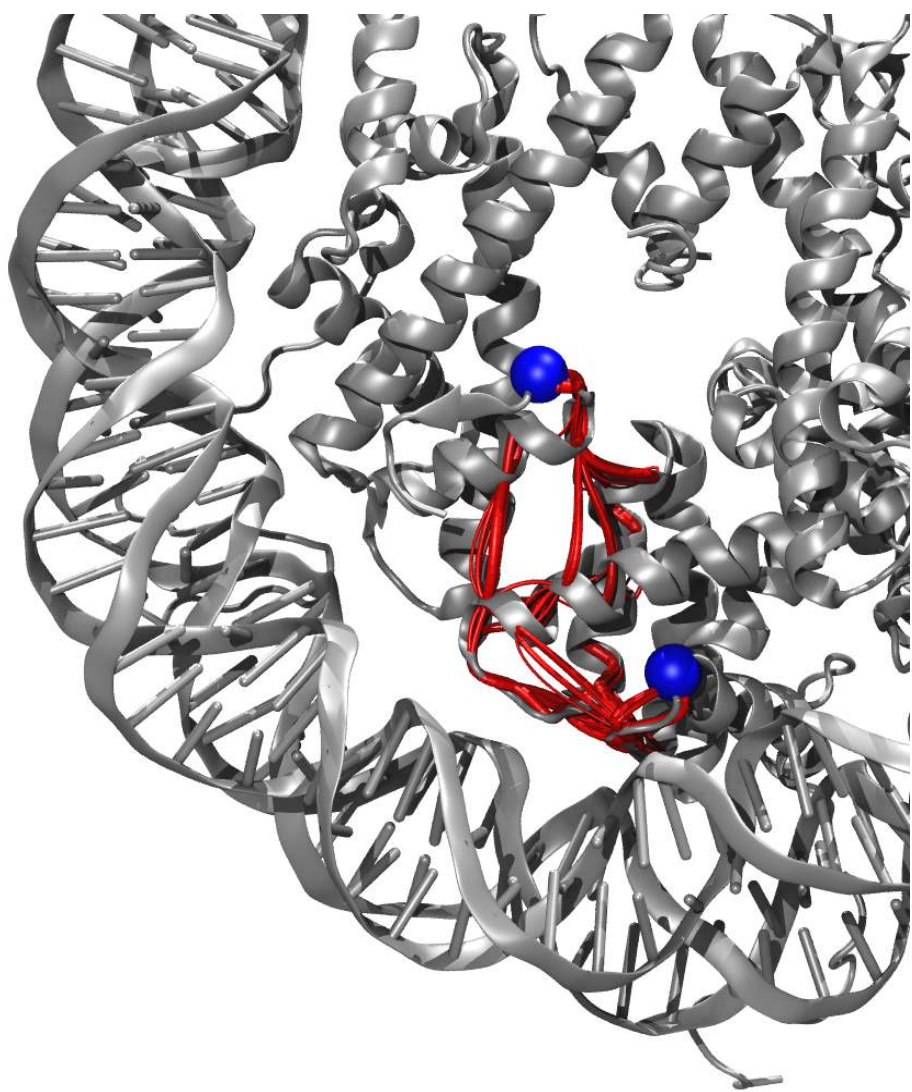

**Figure S21:** Sub-optimal pathways for L1 loop to non-associated docking domain in the L1-Mutant nucleosome. The opposing dimer has been removed to improve visualization clarity.

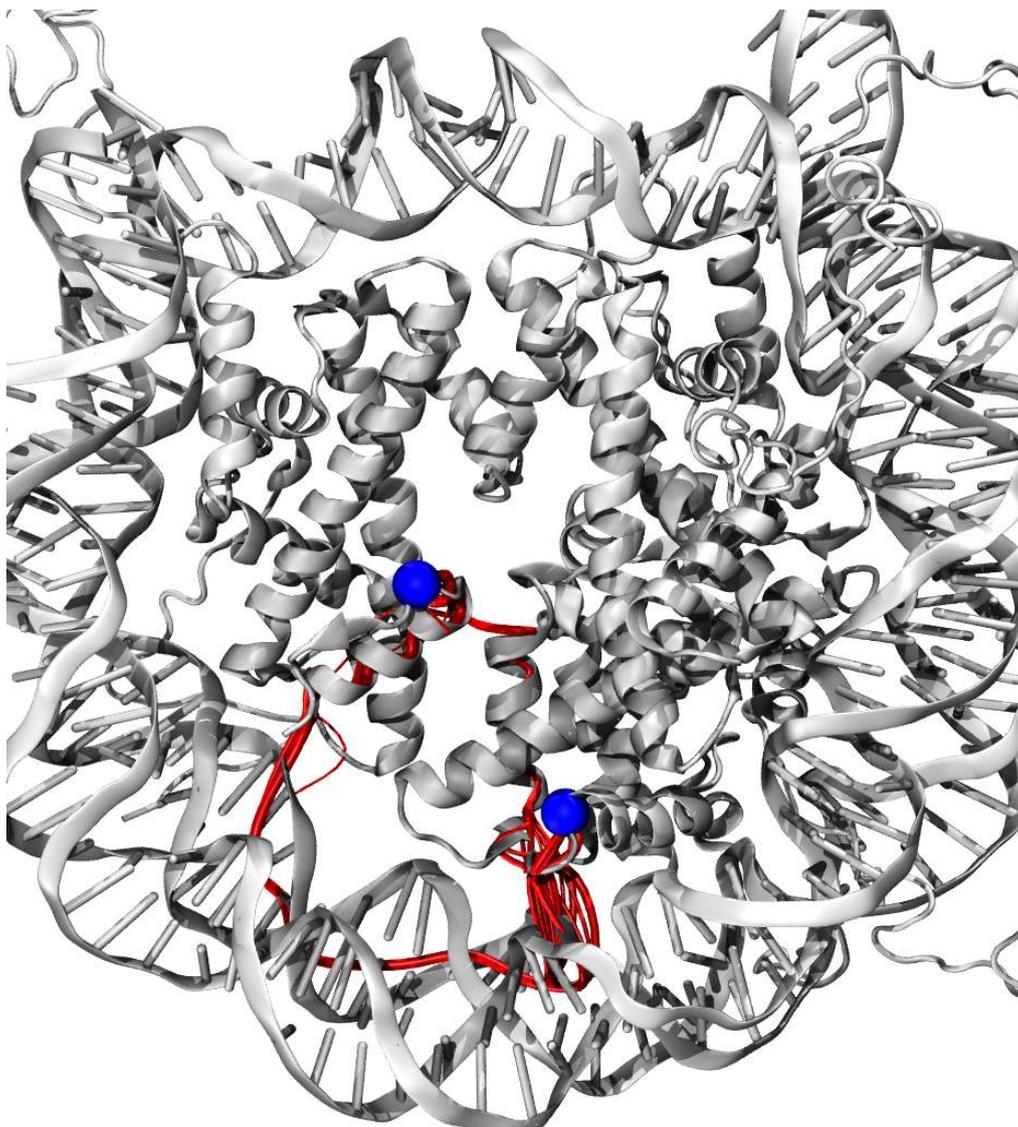

**Figure S22:** Sub-optimal pathways for L1 loop to non-associated docking domain in the macroH2A nucleosome. The opposing dimer has been removed to improve visualization clarity.

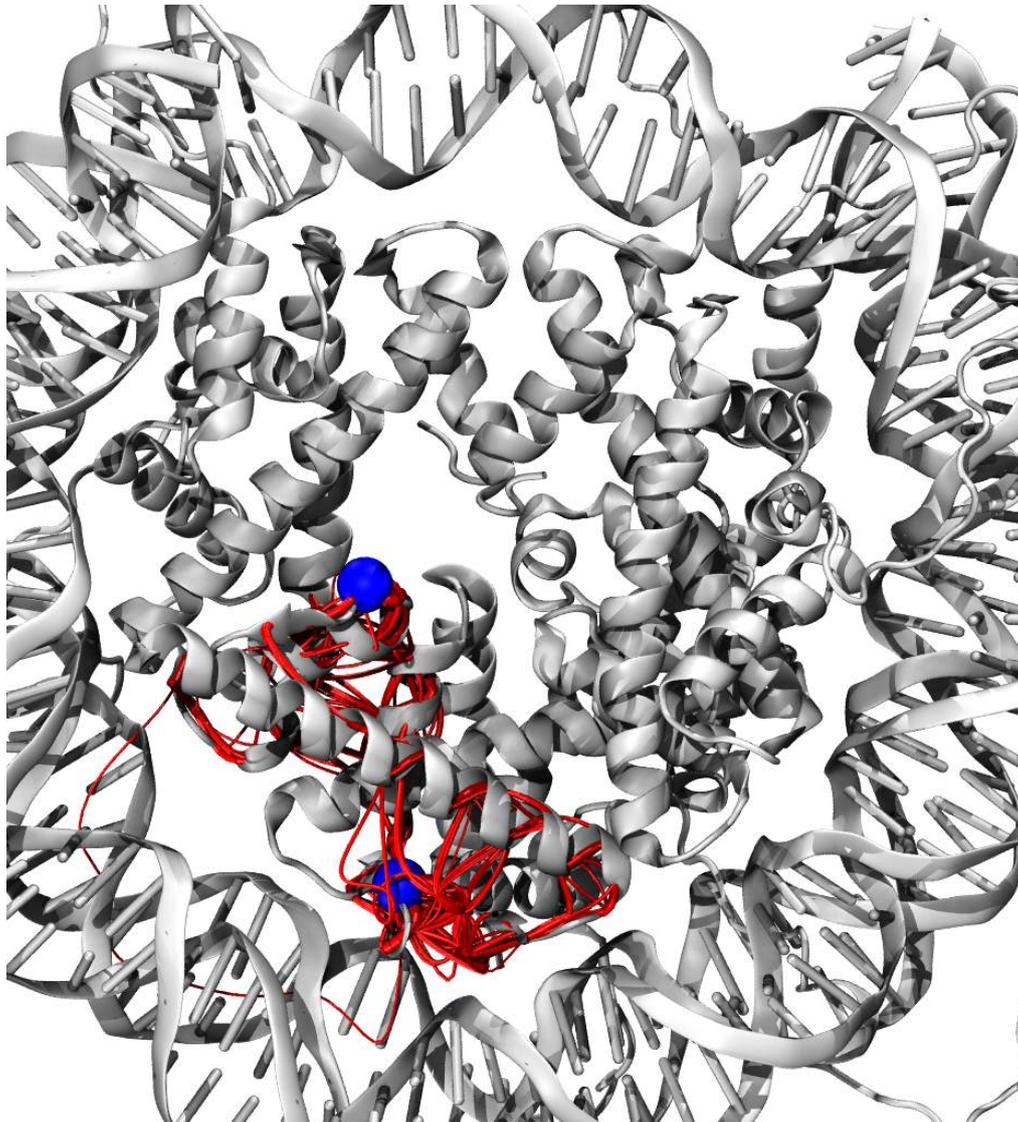

**Figure S23:** Sub-optimal pathways for L1 loop to non-associated docking domain in the H2A.Z nucleosome. The opposing dimer has been reduced to only H2B α2 to improve visualization clarity.